# Identification of possible differences in coding and non coding fragments of DNA sequences by using the method of the Recurrence Quantification Analysis

Sergio Conte

*School of Advanced International Studies on Theoretical and non Linear Methodologies in Physics. Bari-Italy*

and

Alessandro Giuliani

*Istituto Superiore di Sanità, Viale Regina Elena 299, 0016, Rome, Italy*

**Abstract** : *Starting with the results of Li et al. in 1992 there is valuable interest in finding long range correlations in DNA sequences since it raises questions about the role of introns and intron-containing genes. In the present paper we studied two sequences that are the human T-cell receptor alpha/delta locus, Gen-Bank name HUMTCRADCV, a non-coding chromosomal fragment of M = 97630 bases (composed of less than 10% of coding regions), and the Escherichia Coli K12, Gen-Bank name ECO110K, a genomic fragment with M = 111401 bases consisting of mostly coding regions and containing more that 80% of coding regions. We attributed the value (+1) to the purines and the value (-1) to the pirimidines and to such reconstructed random walk we applied the method of the Recurrence Quantification Analysis(RQA) that was introduced by Zbilut and Webber in 1994. By using dimension D=1 and Embedded Dimensions D=3 and D=5, we obtain some indicative results. Also by a simple eye examination of the reconstructed maps, the differences between coding and non coding regions are evident and impressive and consist in the presence in non-coding regions of long patches of the same colour that are absent in the coding sequence. At first sight this suggests a simple explanation to the concept of 'long-range' correlation. On the quantitative plane, we used the %Rec., the %Det., the Ratio, the Entropy, the %Lam., and the Lmax that, as explained in detail in the text, represent the basic variables of RQA. The significant result that we have here is that both Lmax and Laminarity exhibit very large values in HUMTCRADCV and actually different in values respect to ECO110K where such variables assume more modest values. Therefore we suggest that there is the observed difference between HUMTCRADCV and ECO110K. The claimed higher long-range correlations of introns respect to exons from many authors may be explained here in reason of such found higher values of Lmax and of Laminarity in HUMTCRADCV respect to ECO110K*

## 1. Introduction

Rather recently, there has been some interest in the finding of long range correlations in DNA sequences. By using spectral analyses, Li *et al.* found [1] that the frequency spectrum of a DNA sequence containing mostly introns shows $1/f^{\alpha}$ behaviour, which evidences the presence of long range correlations. Peng *et al.* [2] used a random walk model that also led to the conclusion that intron-containing DNA sequences exhibit long range correlations, whereas such a correlation was not found for any of the intronless or cDNA segments. These results raised interesting questions about the role of introns and intron-containing genes. However, other investigators questioned whether the previous observations should be considered artefacts linked to the method used by Peng *et al*. Chatzidimitriou-Dreismann and Larhammar [3], on the other hand, made a careful analysis of the same data set and concluded that both intron-containing and intronless DNA sequences exhibit long range correlations. A subsequent work by Prabhu and Claverie [4] also substantially confirmed such results. Alternatively, Voss [5] based his study on equal-symbol correlation, and he showed a power law behaviour for the sequences studied regardless of the percent of intron contents. Other conflicting results were also obtained. Therefore, it remains still an open question whether the long range correlation properties are different for intronless and intron-containing coding regions. This question is particularly cogent even because all the previous mentioned studies rely on a purely statistical, coarse-grain, perspective with only a limited if not null interest in the actual pattern of periodicities along the sequence. DNA sequence can be equated to a 'palinsesto', one of those old books in which different writings and re-writings, errors and repetitions go together one upon the other, and not as the output of a dynamical system with a given generating function. In this respect the discovery of the distribution, length, and specific features of repetitions and periodic patterns is much more relevant than the discovery of a global probability function.

Under this respect the comparison between coding and non-coding regions in terms of sequence periodicities can be of importance, given the global probability distribution approaches gave largely inconclusive results. On the biological side the rising interest in non coding regions coming from the recognition of regulatory roles [6] for these regions is concentrated on the location and character of 'words' i.e. of specific nucleotide patterns along the genome [7].

The aim of the present study is devoted to a possible identification of differences between coding and non coding fragments of DNA sequences by using the method of the Recurrence Quantification Analysis (RQA) that is explicitly based on the location and quantification of specific patches (words) along a time (or spatial) series without any explicit reference to any functional hypothesis.

We give both a proof-of-concept of the superiority of RQA with respect to more formal probabilistic methods to exploit the character of the differences between coding and non-coding regions and an explanation of the syntactic basis of the observed long-range correlations.

## 2. The RQA Methodology

The recurrence plot is the visualization of a square recurrence matrix of distance elements within a cutoff limit. Given a time series $x_P = [P_1, P_2, ..........., P_M]$ of M points $P_i$, first of all a reconstruction of such given time series must be realized in phase space. It is performed by defining time-delayed vectors $V_i$ of the M points $P_i$ that are delayed or offset in time $\tau$. Each point represents a single amplitude (scalar) at a specific instance in time and we have a D-dimensional vector

Vi = Pi + Pi+$\tau$ + Pi+2 $\tau$ + … + Pi+(D-1) $\tau$

Such reconstruction in the so called phase space needs to know into what dimension the dynamic that we aim to explore, is projected. This is the embedding dimension (D). To reach this objective, we need to pick an appropriate delay $\tau$ between sequential time points of the signal. It must be outlined that the selections of embedding dimension and delay are not trivial, but are based on non linear dynamical theory.

*Embedding dimension* (Embed. Dim.) in principle may be estimated by the nearest-neighbor methodology of Kennel, Brown, and Abarbanel [8]. Dimension is increased in integer steps until the recruitment of nearest neighbours of the dynamics under scrutiny becomes unchanging. At this particular value of dimension, the information of the system has been maximized and, technically speaking, the attractor has been completely unfolded. The so called False Nearest Neighbors ( FNN) may be employed .

The *Delay* $\tau$ (or Time Delay ) represents the basic parameter of interest, and it is selected so as to minimize the interaction between points of the measured time series. This, in effect, opens up the attractor (assuming one exists), by presenting its largest profile. Two common ways of selecting a proper delay include finding the first minimum in either the (linear) autocorrelation function or (nonlinear) mutual information function [9] of the given time series.

The further step is to compute the distances between all possible combinations of i-vectors (Vi) and j-vectors (Vj) according to the norming function selected. The minimum or maximum norm at point Pi, Pj is defined, respectively, by the smallest or largest difference between paired points in vector-i and vector-j. The Euclidean norm is defined by the Euclidean distance between paired vectors.

Computed distance values are distributed within a distance matrix DM[j, i] which aligns vertically with vector time series (T) of N scalar points P, Tj, and horizontally with Ti (i = 1 to W; j = 1 to W; where maximum W = N – M + 1). The distance matrix has $W^2$ elements with a long central diagonal of W distances all equal to the pair (0, 0). This ubiquitous diagonal feature arises because individual vectors are always identical matches with themselves (Vi = Vj whenever i = j). The matrix is also symmetrical across the diagonal since that if vector Vi is close to vector Vj then also vector Vj is close to vector Vi.

The first recurrence parameter to be considered, is the *radius*. Briefly, all (i, j) elements in recurrence matrix with distances at or below the Radius cutoff are included in the recurrence matrix (element value = 1), but all other elements are excluded from recurrence matrix (element value = 0). In detail, recurrence plots, especially coloured versions, express recurrence distances as contour maps.

Recalling the brief history of recurrence analysis, we recall here that the recurrence plots were originally introduced as qualitative tools to detect hidden rhythms graphically [10]. The next step was to promote recurrence analysis to quantitative status. Zbilut and Webber introduced RQA, the Recurrence Quantification Analysis [11].

The first recurrence variable of RQA is *%recurrence* (%Rec). %Rec quantifies the percentage of recurrent points falling within the specified radius. This variable can range from 0% (no recurrent points) to 100% (all points recurrent). It identifies recurrent or rather periodic patterns in the given time series.

The second recurrence variable is *%determinism* (%Det).

%DET measures the proportion of recurrent points forming diagonal line structures. Diagonal line segments must have a minimum length defined by the line parameter. The name determinism comes from repeating or deterministic patterns in the dynamic. Periodic signals (e.g. sine waves) will give very long diagonal lines, chaotic signals will give very short diagonal lines, and stochastic signals (e.g. random numbers) will give no diagonal lines at all (unless parameter Radius is set too high).

The third recurrence variable is *linemax* (LMax), which is simply the length of the longest diagonal line segment in the plot, excluding the main diagonal line of identity (i = j). This is a very important recurrence variable because it inversely scales with the most positive Lyapunov exponent [11]. Positive Lyapunov exponents gauge the rate at which trajectories diverge, and are the hallmark for dynamic chaos. Thus, the shorter the linemax, the more chaotic (less stable) is the signal, [12].

The fourth recurrence variable is *entropy* (Ent), which is the Shannon information entropy of all diagonal line lengths distributed over integer bins in a histogram. Ent is a measure of signal complexity and is calibrated in units of bits/bin.

The fifth recurrence variable is *trend* (Tnd), which quantifies the degree of system stationarity. If recurrent points are homogeneously distributed across the recurrence plot, TND values will hover near zero units. If recurrent points are heterogeneously distributed across the recurrence plot, Tnd values will deviate from zero units.

The sixth and seventh recurrence variables, *%laminarity* (%Lam) and *trapping time* (TT), were introduced by Marwan, Wessel, Meyerfeldt, Schirdewan, and Kurths [13]. %Lam is analogous to %Det except that it measures the percentage of recurrent points comprising vertical line structures rather than diagonal line structures. The line parameter still governs the minimum length of vertical lines to be included. TT, on the other hand, is simply the average length of vertical line structures. Laminarity, quantifying vertical structures, identifies transitions as chaos-chaos, chaos to periodic, thus intermittency, unstable singularities [14].

It remains to outline that in the aim of cross correlation, it is possible to detect recurrence patterns shared by paired signals by cross recurrence analysis (KRQA) [13,15]. The mathematics of cross recurrence, as well as the parameters and variables of cross recurrence, all are the same as explained for auto-recurrence. Synchronization or coupling between two signals enforces %Rec and %Det and lowers Entropy values in KRQA respect to RQA. The viceversa happens for two considered uncoupled signals.

Finally, we outline that RQA may also be applied by selecting sub-matrices of the original recurrence matrix. This is to say that we calculate the previously discussed RQA variables by epochs.

Let us delineate now the manner in which RQA may be applied in analysis of DNA sequences.

Let us start with an example that was originally given by Zbilut and Webber. Let us ask how it is that we may write a text containing thousands of words since we use only 22 alphabet letters, writing as example in Italian. The obvious answer is that the letters must be reused. They show recurrences. So at the word level or orthographic (spelling) level, symbols are simply reused in any combination desired by the author, as long as they correspond to allowable words in the language of choice. Common experience informs us that letters in words or words in sentences do not, and need not, appear at periodic intervals. Rather, actual linguistic sequences are at once highly nonlinear and highly meaningful as well. In this context, Orsucci, Walter, Giuliani, Webber, and Zbilut [16] implemented RQA to study the linguistic structuring of American poems, Swedish poems, and Italian translations of the Swedish poems.

They found invariance among the various language exemplars, suggesting hidden structuring at the orthographic level. It is intriguing to consider the potential ability of recurrence strategies in the analysis of written text or spoken words as first explored by Orsucci et al. [16]. We may proceed at the orthographic level, rendering any speech text numeric by arbitrarily substituting integers for letters: A=1; B=2; C=3;…; X=24; y=25; Z=26; and for numbers: 0=27; 1=28; 2=29 …; 7=34; 8=35; 9=36. There is, however, an important and determinant feature that we must outline here. In using RQA in this case we must put our attention on the manner in which the recurrence parameters of RQA must be set. Since the encoding scheme is entirely arbitrary (we could have used: Z=1; Y=2; X=3; …; etc.), the most important constraint is that the Radius, previously discussed in our RQA exposition, now must be set to 0 distance units. This will insure that only identical letters (unique integers) will recur with each other. The embedding dimension can be set to one or higher, but for M > 1 the delay should be set to one so as not to skip any letters in the string. With these preliminaries it appears evident that diagonal line structures in the recurrence plot will be of interest. If only identical letters count as recurrent points, a string of diagonal recurrences must indicate that the identical string of characters appears at different positions in the text. Actually, lines of varying length must represent words of varying length. Lmax and %Laminarity also will be of interest. In brief, using all the previously discussed RQA variables, important features of recurrence quantifications can be captured in the given sequence of symbols. Using this criterion we may use RQA to analyse DNA sequences where in this case the alphabet is given by four letters $[A, C, G, T]$. A way is to apply the method of RQA to DNA sequences, as it was made in previous papers as example by Frontali, and Pizzi [17,18], realizing a series by the symbolic codification $A = 1, T = 2, C = 3, G = 4$. In the case of the present paper we will use a different codification. Given the DNA sequences, as often it was made also by other investigators, we attribute the value (+1) to the purines and the value (-1) to the pirimidines. We realize in this manner a kind of random walk based on the binary values ($\pm 1$), this character of 'random walk' makes the chosen code similar in spirit to the well know Chaos Game Representation (CGR) widely used in genome scale sequence comparisons while allowing for a much more precise location of the detected periodicities [19].

Incidentally, we observe here that rather recently a reconstruction of genetic code in the dyadic plane was obtained [20]

**3. The Selected Nucleotide Sequences.**

We analysed the two DNA sequences studied in Ref. [21]. These two sequences are the human T-cell receptor alpha/delta locus, Gen-Bank name HUMTCRADCV, a non-coding chromosomal fragment of $M = 97630$ bases (composed of less than 10% of coding regions), and the Escherichia Coli K12, Gen-Bank name ECO110K, a genomic fragment with $M = 111401$ bases consisting of mostly coding regions and containing more that 80% of coding regions. The choice of these two sequences was dictated by the fact the authors detected only a relatively minor difference between coding and non coding distances in the realm of the same distribution model, thus representing a potentially useful work-bench for a local model like the one we propose.

## 4. Results of RQA

Let us repeat that in the RQA we used a Radius, $R = 0$, as previously discussed, so to insure that only identical symbols of our sequences will be estimated to recur with each other. As previously outlined, we used a time delay $\tau = 1$, and we selected a line $L = 3$. We considered embedding dimension D=1 (only one symbol of the sequence, that is a single nucleotide with A or G equal to +1 and C or T equal to -1), D=3 (three symbols for each point. A sequence of three nucleotides, A or G equal to +1 and C or T equal to –1 in our reconstructed random walk, a), and, finally, D=5 (five symbols for each point. A sequence of five nucleotides, A or G equal to +1 and C or T equal to –1 in our reconstructed random walk).

Let us look first at the obtained recurrence plots. Figures 1a, 1b, 1c, 1d, 1e, 1f, 1g give the results for non-coding HUMTCRADCV in the case D=1, D=3, D=5, respectively. Figures 2a, 2b, 2c give the same recurrence plots in the case of the coding ECO110K.

Looking at the Fig.1a one immediately appreciates the whole map of the examined sequence. Only two values of distances are possible, respectively Euclidean distance=0, corresponding to repeat of the same sequence and signed in the plot by white, and Euclidean distance=2 corresponding to different sequences, and signed in the plot by dark colour. Any detail of the mapped sequence may be evidenced. As example in Fig. 1b we have a section of the same map.

In Fig. 1c we have a further selected section of the whole map, and we may appreciate the presence of diagonal as well as well vertical lines corresponding to Determinism and Laminarity, respectively. As example let us look at the pair of points (50351, 67268), (50352, 67269), (50353, 67270) or (50337, 67278), (50337, 67277), (50337, 67276).

In Fig. 1d, we have the recurrence plot of the same non-coding HUMTCRADCV in the case of Embed. Dim. D=3. In Fig. 1e it is given instead a selected section of this map where again we may identify the basic features of the RQA variables.

In Figures 1f and 1g we have the recurrence plot of HUMTCRADCV and a selected section of the whole map for Embed. Dim. =5, respectively.

In Fig. 2a we give the recurrence plot in the case of coding ECO110K, Embed. Dim. =1 while in Fig. 2b we have a selected its section. In Fig. 2c we have the recurrence plot of coding ECO110K in the case of embedding Dimension D=3 with selected section given in Fig. 2d, and, finally, in Fig. 2e we have the case of Embedding Dimension D=5 with selected section given in Fig. 2f.

All the elaborations were performed by using a commercially available version called VRA (Visual Recurrence Analysis) that can be obtained at http://home.netcom.com/~eugenek/download.html.).

By a simple eye examination the texture differences between coding and non coding regions are evident and impressive and consist in the presence in non-coding regions of long patches of the same colour that are absent in the coding sequence. At first sight this suggests a simple explanation to the concept of 'long-range' correlation, but we need to go quantitative.

We may now examine the results by using RQD software so to get the quantitative values of the above mentioned RQA descriptors.

Concerning Recurrence Quantification Analysis, the original programs that we used, were developed by Webber and Zbilut in 1994, and can be downloaded at http://homepages.luc.edu/~cwebber., whereas a MATLAB version of RQA, developed at the University of Postdam called CRP toolbox can be found at http://tocsy.agnld.uni-postdam.de developed by Marwan *et al*., in 2007).

We used RQD to calculate all the previously introduced RQA variables for both the examined DNA sequences in order to estimate the %Rec, the %Det, the Lmax, the Entropy, the % Laminarity, and the Ratio (%Det/%Rec) for the whole DNA sequences that we took in consideration.

Let us resume briefly the meaning of such variables for given time series.

%Rec expresses the percent ratio between the number of recurrent points and the total number of points. For % Determinism it is important to outline again that processes with uncorrelated or weakly correlated behaviour cause none or very short diagonals, whereas deterministic processes cause longer diagonals. The Lmax relates the longest diagonal line found in the RP, and its inverse, the divergence (DIV), measures the divergence of the series calculated as Div= 1/Lmax.

These measures are related to the exponential divergence of the phase space trajectory. The faster the trajectory segments diverge, the shorter are the diagonal lines and the higher is DIV. Still, Entropy, Entr., reflects the complexity of the RP in respect of the diagonal lines. For uncorrelated noise the value of Entr. is rather small, indicating its low complexity.

We can find vertical lines in presence of laminar states in intermittence regimes. Laminarity represents the occurrence of laminar states in the system without describing the length of these laminar phases. Lam will decrease if the RP consists of more single recurrence points than vertical structures. In contrast to the RQA measures based on diagonal lines, these measures are able to find chaos-chaos, chaos –order transitions, intermittency, singularities often evaluated also in relation with LMax.

The results of our investigation are reported in Table1 for embedding dimensions D=1, D=3, D=5, respectively. Let us remember that the aim of our investigation is to ascertain differences between coding and non-coding regions of the considered fragments. Looking at the results of Table 1, we deduce that such two fragments do not exhibit net differences between coding and on-coding fragments for the case D=1, D=3, D=5 for the RQA variables %Rec, %Det, Entropy and Ratio.

Net differences result instead for two RQA variables of basic importance. They are the Lmax and the % Laminarity. In brief, the two coding and non-coding fragments taken in consideration, seem to evidence a very similar tendency for their recurrent behaviour (periodicity in the repeat of nucleotidic symbolic representation), a very similar level of determinism, still a very similar level of complexity as characterized by the Entropy, but they show net and substantial differences in the values of Lmax and % Laminarity. May be of importance to outline here that this tendency in their behaviour is confirmed as well as when we examine the embedding dimension D=1 as in the case D=3 and in D=5.

The implications to have found significant differences in Lmax and in % Laminarity of coding respect to non-coding DNA sequences is of interest. Therefore, before of its definitive acceptance, it is necessary to perform some further control. On the general plane, the method of surrogate data, see for example Schreiber and Schmitz [22] for a review, has become a useful tool to address the question if the irregularity of the data is most likely due to non linear deterministic structure or rather due to random inputs to the system or fluctuations in the parameters.

The method of surrogate data, which was first introduced by Theiler *et al.* [23] in non linear time series analysis, consists of generating an ensemble of "surrogate" data sets similar to the original time series, but consistent with the null hypothesis, usually that the data have been created by a stationary Gaussian linear process. The subsequent step is of computing a discriminating statistic for the original and for each of the surrogate data sets. We can create surrogate data by taking their fast (discrete) Fourier transform (FFT) and multiplying it by a random phase parameter uniformly distributed in $(0, 2\pi)$, then it is possible to compute the inverse of FFT and we have a time series with the prescribed spectrum. Different realization of the random phase gives new surrogate data. This process of phase randomisation preserves the Gaussian distribution.

Another way is that the data values are simply shuffled, and we used here this second technique. The results are reported in Table 2.

By inspection of the results given in Table 2 one sees that actually do not result significant differences for all the RQA variables in the cases of original respect to shuffled data. However, the only significant differences appear for Lmax and % Laminarity and such differences result more marked in the case of non coding HUMTCRADCV respect to ECO110K-genomic fragment. In conclusion, such two RQA variables represent the most interesting reference in the kind of analysis that we are developing.

The next step of our analysis was based on RQE. As previously discussed, generally speaking such kind of RQA analysis consists in introducing windows in the given time series and thus estimating the RQA variables in such prefixed epochs. We selected epochs of 1000 points, that is to say, we analysed sequences of 1000 nucleotides (A or G symbolized by +1 and C or T symbolized by –1) in the proper embedding dimension previously explored, D=1, D=3, D=5. Time delay again was considered $\tau = 1$. The data shift was posed equal to 1000. Again the Radius R was posed equal to zero and the Line was considered equal to 3. In the case of non coding HUMTCRADCV, we examined a total of 97 epochs, while in the case of ECO110K-genomic fragment we had a total of 111 epochs. As final step of our analysis we used the KRQE. As we discussed in the previous section, the KRQE states for cross recurrence analysis by epochs, and, generally speaking, in time series it is useful to identify synchronization between the two given series or, more generally, existing coupling and functional interdependence. We applied KRQE between non coding HUMTCRADCV and ECO110K-genomic fragment with the aim to ascertain possibly the existence of some kind of interdependence between such two nucleotidic sequences. Again we used epochs of 1000, data shift equal to 1000, embedding dimension, D=1, D=3, D=5, time delay $\tau = 1$, Radius =0, and line L=3. We calculated also the mean values obtained for each considered RQA variable.

The results are given in Figures 3a-c, 4a-c, 5a-c, 6a-c, 7a-c, 8a-c for RQA variables %Rec, %Det, Entropy, Ratio, Lmax, and %Laminarity.

Let us start with the results relative to %Rec. Looking at the Fig.3a we have the results in the case D=1.

First let us look at the numerical values. It is seen that both non coding HUMTCRADCV and ECO110K-genomic fragment have the almost equal mean values that also return in the case of KRQE.

Therefore, HUMTCRADCV and ECO110K show the same mean value of recurrences, respectively in the 97 and 111 epochs taken in consideration. Under such view points the two given sequences have not significant differences. Looking at the Figure 3a we see that the profile of %Rec in the two cases of HUMTCRADCV respect to ECO110K seems to evidence some difference in the sense that the tendency in HUMTCRADCV seems to exhibit more peaked values in some epochs respect to ECO110K. The conclusion seems to be that HUMTCRADCV and ECO110K do not show significant differences in recurrences also if, considering their behaviour, it seems that HUMTCRADCV has a resulting behaviour that in some epochs results lightly more recurrent respect to ECO110K. There is still to observe that the obtained numerical results run about 50% for %Rec that is a unusual very high value in RQA analysis. Usually, numerical values about 2-15% are considered in RQA in order to avoid the influence of noise. Of course a bad evaluation of %Rec inevitably affects soon after the estimated values of %Det. It is usually obtained for very high selected values of the Radius.

Let us now examine the case of %Rec in the case of Embedding Dimension D=3. It is given in Fig. 3b.

Concerning the obtained numerical values it is seen that %Rec remains substantially similar in HUMTCRADCV and ECO110K also if a light more evident prevalence of recurrences now appears in HUMTCRADCV respect to ECO110K in the case D=3 respect to D=1. Looking at the graphical behaviour of the variable by epochs, one see that such two profiles show some differences in the sense that % Rec. peaks in HUMTCRADCV appear always higher and more defined respect to ECO110K. Also the mean values of %Rec remain here about 13% that represents a more acceptable value respect to the previous case obtained for D=1. The lower values obtained by KRQE confirm that we have not interdependence or coupling between HUMTCRADCV and ECO110K. In conclusion, also in the case D=3, we have very similar values for %Rec. in HUMTCRADCV and ECO110K with possibly a very light prevalence in HUMTCRADCV respect to ECO110K.

Let us now examine the case of Embedding Dimension D=5. The results are given in Fig. 3c.

It is very similar to the case of D=3 and therefore it will not be discussed further here. The only required observation is that the data of %Rec. run now about 3-6% that represent very reasonable results.

We may now pass to introduce the results obtained in % Determinism.

Let us start again by the case of Embedding Dimension D=1. The results are given in Fig. 4a.

The obtained mean values indicate that we have a modest but not significant prevalence of %Det. in non-coding HUMTCRADCV respect to ECO110K-genomic fragment. The graphic behaviour by epochs confirms such tendency. The very low values of KRQE still confirm that we have not coupling or interdependence between HUMTCRADCV and ECO110K.

The same tendency is confirmed in the case of Embedding Dimension D=3 and D=5 as reported respectively in Figures 4b and 4c.

The differences between HUMTCRADCV (52-53% in HUMTCRADCV against 50-51% in and ECO110K) remains rather modest. However, this is a tendency that results constantly confirmed in all the embedded cases.

A speed look to the results obtained for Entropy, given in Figures 5a-c, and to the Ratio, given in Figures 6a-c, confirm that we have not marked differences also for such estimated variables in
HUMTCRADCV respect to ECO110K.

It remains to give now the results about the last two RQA variables, the Lmax that is given in Figures 7a-c, and the %Laminarity, given in Figures 8a-c.

In relation to the calculated mean values in the case of Embed Dim D=1, we observe that a marked difference arises in HUMTCRADCV respect to ECO110K. Also the behaviour of this variable by epochs results to be different in HUMTCRADCV respect to ECO110K.

The same tendency is observed in the case of embedding dimension D=3 as well as in the case D=5.

Such result is supported also from the previous analysis performed by RQE where in fact we obtained an LMax value of 61 against 26 for HUMTCRADCV respect to ECO110K in the case of D=1, of 59 against 24 in the case D=3, and of 57 against 22 in the case D=5. Observing the behaviour by epochs of Lmax in the cases D=1, D=3, and D=5 we re-find epochs in HUMTCRADCV with a number of peaks with very great values of Lmax, ranging about 40-50 respect to ECO110K that quite systematically exhibit lower values of such peaks.

Let us examine now the last RQA variable, the %Laminarity. The results are given in Figures 8a-c

There is no doubt in this case that we are in presence of a net difference of non coding HUMTCRADCV respect to ECO110K-genomic fragment. In accord also with the previous results obtained by RQE and reported in Table 1, looking at the values here obtained for mean values, we conclude that %Laminarity evidences with systematic behaviour greater values in HUMTCRADCV respect to ECO110K. Also the datum obtained by the graph of %Lam by epochs confirms this systematic tendency.

The same tendency is obtained also when we consider the results in the case of Embed. Dimension D=3 and D=5. Observing the results of Figures 8b and 8c, we conclude without any doubt that % Laminarity discriminates in its values and in its behaviour the HUMTCRADCV fragment respect to ECO110K.

In conclusion, summarizing our RQA analysis, we may affirm that the two examined sequences (non coding HUMTCRADCV and ECO110K-genomic fragment) exhibit some interesting differences. They have been obtained first of all by the RQD and confirmed soon after by RQE and KRQE analysis. Very light differences have been found in %Rec and %Det variables in HUMTCRADCV respect to ECO110K but constantly expressed in favour of HUMTCRADCV respect to ECO110K. Substantial difference have been found by the Lmax. Finally, a net discrimination between HUMTCRADCV respect to ECO110K has been obtained by %Laminarity. Therefore, %Lam and Lmax represent the two RQA variables constituting the most important object of our investigation. In this framework we have completed our analysis with Figures 9a-c and 10a-c where %Lam against Lmax are represented.

**5. Discussion**

In the present paper we have examined two sequences corresponding to: 1) the human T-cell receptor alpha/delta locus, Gen-Bank name HUMTCRADCV, a non-coding chromosomal fragment of $M = 97630$ bases (composed of less than 10% of coding regions), and 2) the Escherichia Coli K12, Gen-Bank name ECO110K, a genomic fragment with $M = 111401$ bases consisting of mostly coding regions (around 80%). We have employed the method of the Recurrence Quantification Analysis (RQA), and in detail we have used the RQD, the RQE and the KRQE. In addition to the dimension D=1, an embedding dimension D=3 and D=5 have been considered. The sequences data have been also shuffled as we explained previously in order to ascertain the actual differences between the given and shuffled data.

The results of the RQD analysis have been given in Table 1. They evidence that the HUMTCRADCV has a very moderate and not significant increase of % Rec, of %Det, Entropy, and Ratio respect to ECO110K while instead a net and significant difference is found for Lmax and %Laminarity. We have also performed the RQE analysis considering sequences of 1000 symbols shifted of 1000 by each epoch. The results, previously obtained by RQD, have been confirmed by RQE analysis. Again it may be noted a modest tendency for HUMTCRADCV respect to ECO110K to prevail for %Rec and % Det without having statistical significance. Instead, a net and significant difference has been found for Lmax and for % Laminarity. By application of the KRQE it has been possible to conclude that there is not any dependence or coupling of one sequence respect to the other. In conclusion, according to the RQA results, two are the variables on which we must concentrate mainly our attention. They are the Lmax and the %Laminarity. In the previous sections we gave the conceptual explanation concerning the meaning of such two variables. Let us help here with the support of an example. Lmax means that we find pieces of sequences that repeat themselves in different positions as AGGGCTCGTTT….. and AGGGTCGTTT…. while Laminarity relates repetitive pieces of sequences as AAAAAAAAAA or GGGGGGGGGG . The important result here is that both Lmax and Laminarity exhibit very large values in HUMTCRADCV and actually different in values respect to ECO110K where such variables assume more modest values. Here there is the observed difference between HUMTCRADCV and ECO110K. The claimed higher long – range correlations of introns respect to exons from many authors may be explained here in reason of such found higher values of Lmax and of Laminarity in HUMTCRADCV respect to ECO110K. In substance HUMTCRADCV always exhibit lightly higher values respect to ECO110K for %Rec and %Det and this line of tendency just indicates that HUMTCRADCV may be moderately more recurrent and more deterministic respect to ECO110K, but the very significant difference consists in the basic fact that HUMTCRADCV contains very large and repetitive pieces of sequences that instead do not appear in ECO110K. Some further consideration deserves the results that we have obtained.

We know that, generally speaking, laminar flow occurs when a fluid flows in parallel layers, with no disruption between the layers. Laminar flow is a flow regime characterized by high momentum diffusion, low momentum convection, pressure, and velocity independent from time. It is the opposite of turbulent flow. This is the reason because such term was introduced in RQA analysis for time series by Marwan when he pointed out that this feature, quantifying vertical structures, identifies transitions such as chaos–chaos, chaos to periodic, and so on. In an excellent paper by Zbilut and Webber it was shown that Laminarity and Lmax are two RQA variables that can demonstrate the presence of unstable singularities which are often found in biological dynamics [14] This is the reason because we introduced the Figures 9a-c and 10a-c in which we represented Lmax vs Laminarity respectively for HUMTCRADCV and ECO110K in the dimensions D=1, 2, 3. By inspection of such figures one observe that there are many epochs in which both Lmax and Laminarity agree in the sense that at the increasing (decreasing) value of one of such two variables corresponds an increasing (decreasing) value for the other variables. However, there are also epochs in which to an increased value in laminarity in one epoch corresponds in the subsequent epoch a corresponding decreased value in Lmax. These should represent singularities as suggested by Zbilut and Webber.

Obviously, in the case of DNA sequences we have to outline that we cannot attribute to terms of unstable singularities the same meaning as it is in the case of time dynamical regimes. In DNA sequences by unstable singularities we must intend , more properly, phases of rearrangement of the sequence, that is to say, generally speaking, cases in which mutations or duplications are possible. In any case we attempted to count the number of agreement of Lmax and % Lam as well as those of possible singularities for HUMTCRADCV and ECO110K, respectively. We find that for HUMTCRADCV the agreements of Lmax with %Lam are about double respect to ECO110K while instead the number of possible singularities in HUMTCRADCV is about an ahlf respect to ECO110K. This last result enforces our result that the observed long-range correlation in non-coding respect to coding sequences is linked to Lmax and % Lam. In particular a non coding sequence seems to exhibit a more stable structure in opposition to the coding sequence that instead points out a more evident tendency to unstability and singularities.

This opens the way to a lot of possible speculations about the regulatory role played by such non-coding sequences and namely to the need to acquire a kind of patterning sufficiently robust to random mutations. In fact, while in coding regions the 'interface with real biology' is given by the resulting protein molecule (that thanks to both the degeneracy of genetic code and the strict control exerted by the entire cell machinery on misfolded intermediates) undergoes a lot of 'quality checks', the non-coding regions play their role (if any, given it seems unreliable that the great majority of non-coding regions is made by junk DNA) as such and thus they have a strong evolutionary constraints in adopting a relative robust scheme.

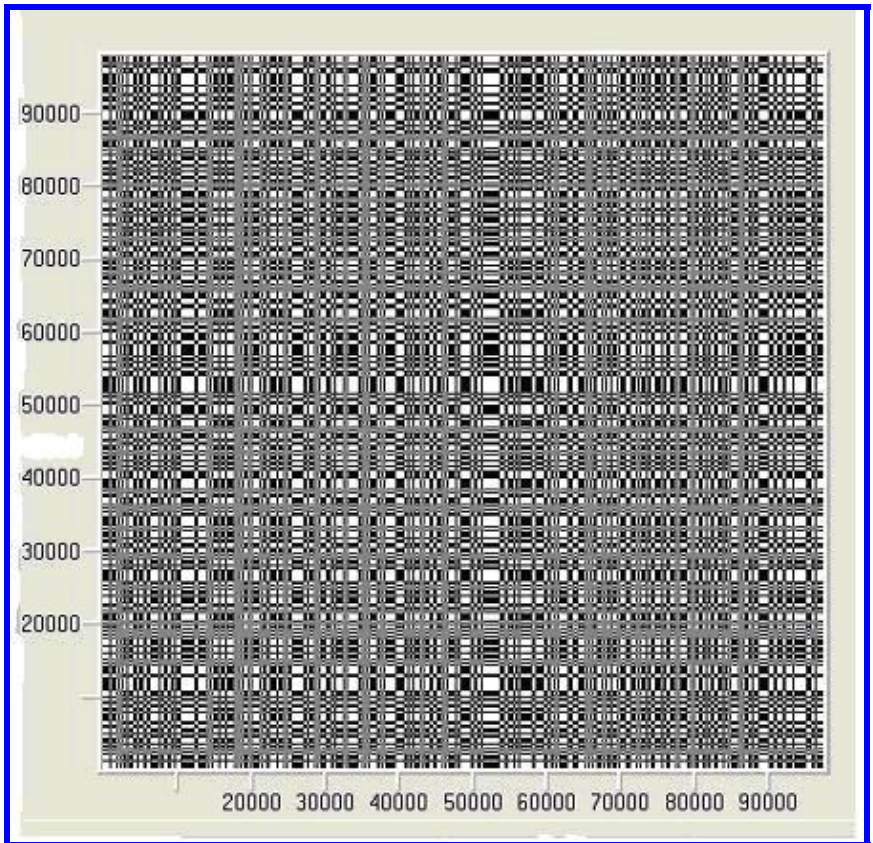
Fig.1a, Recurrence Plot of non-coding HUMTCRADCV (Embed. Dim. D=1)

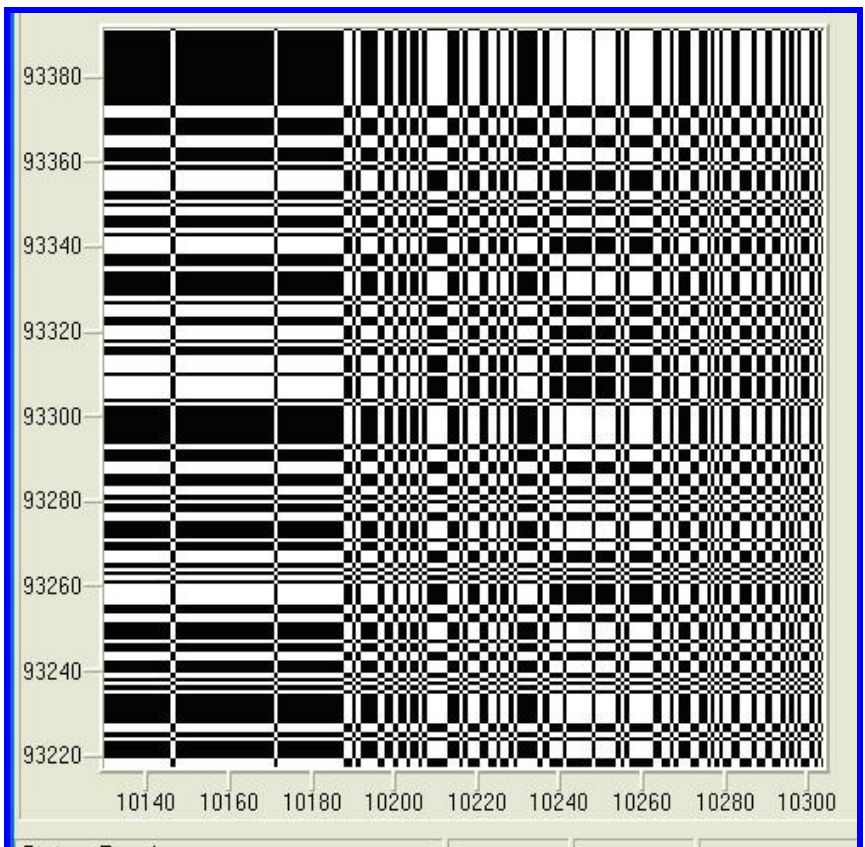
Fig.1b, Recurrence Plot of non-coding HUMTCRADCV (Embed. Dim. D=1)
It is a selected section of the whole map given in Fig.1a

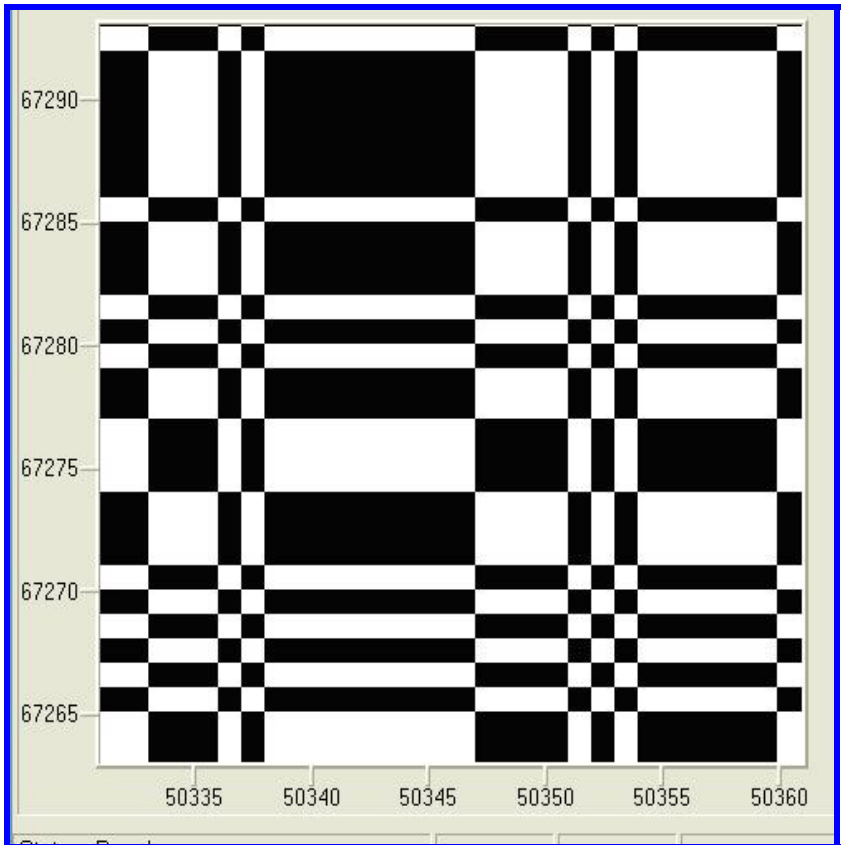
Fig.1c, Recurrence Plot of non-coding HUMTCRADCV (Embed. Dim. D=1)
It is a further selected section of the whole map given in Fig.1a

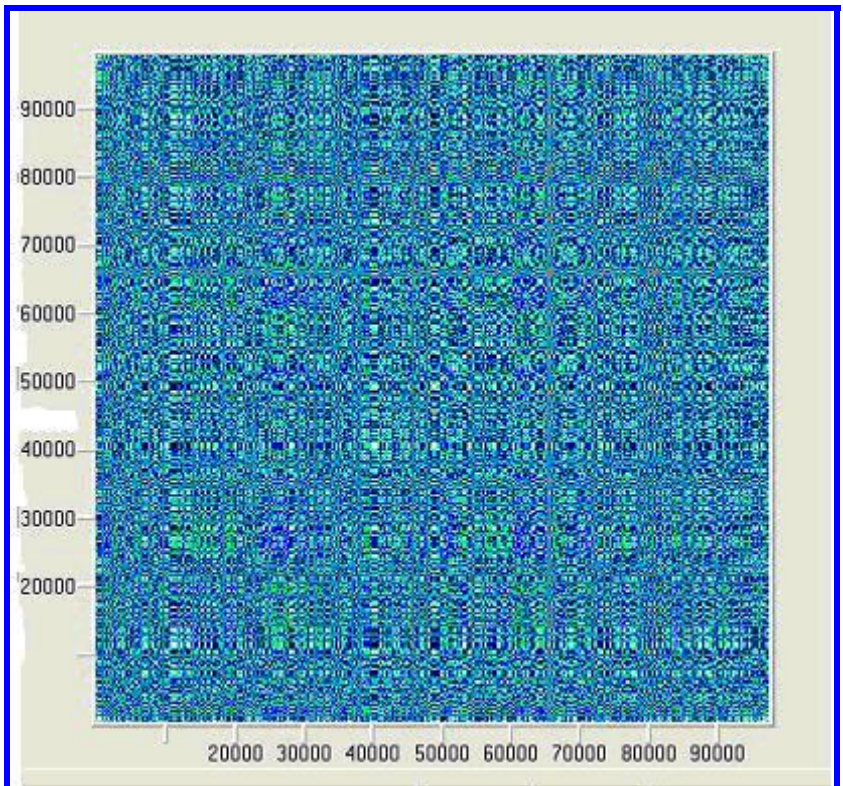
Fig.1d, Recurrence Plot of non-coding HUMTCRADCV (Embed. Dim. D=3)

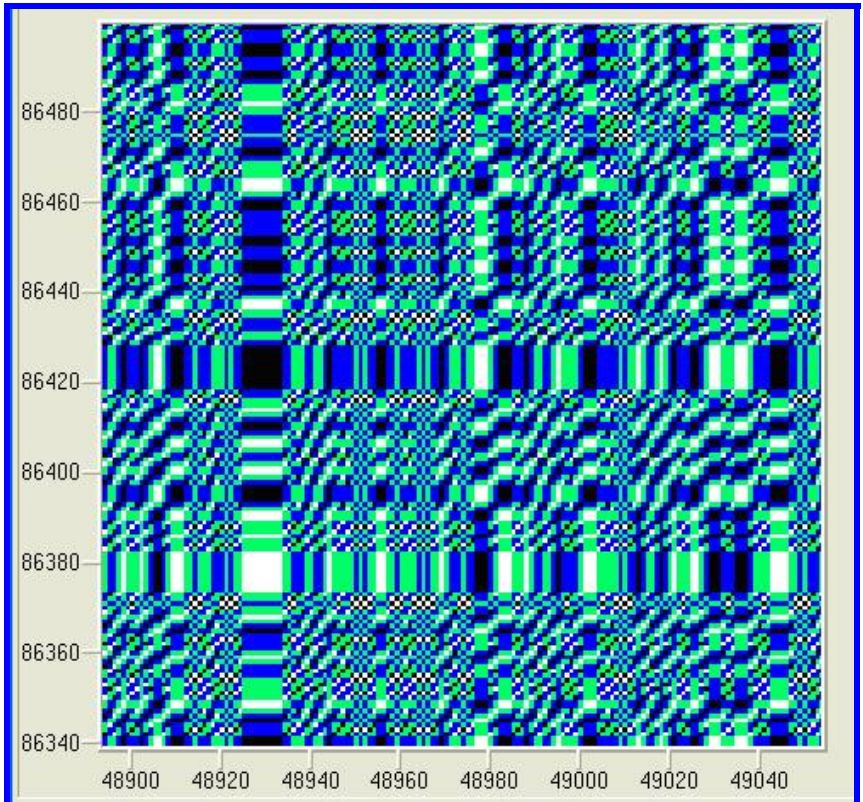

Fig.1e. Recurrence Plot of non-coding HUMTCRADCV (Embed. Dim. D=3)
It is a selected section of the whole map given in Fig.1d.

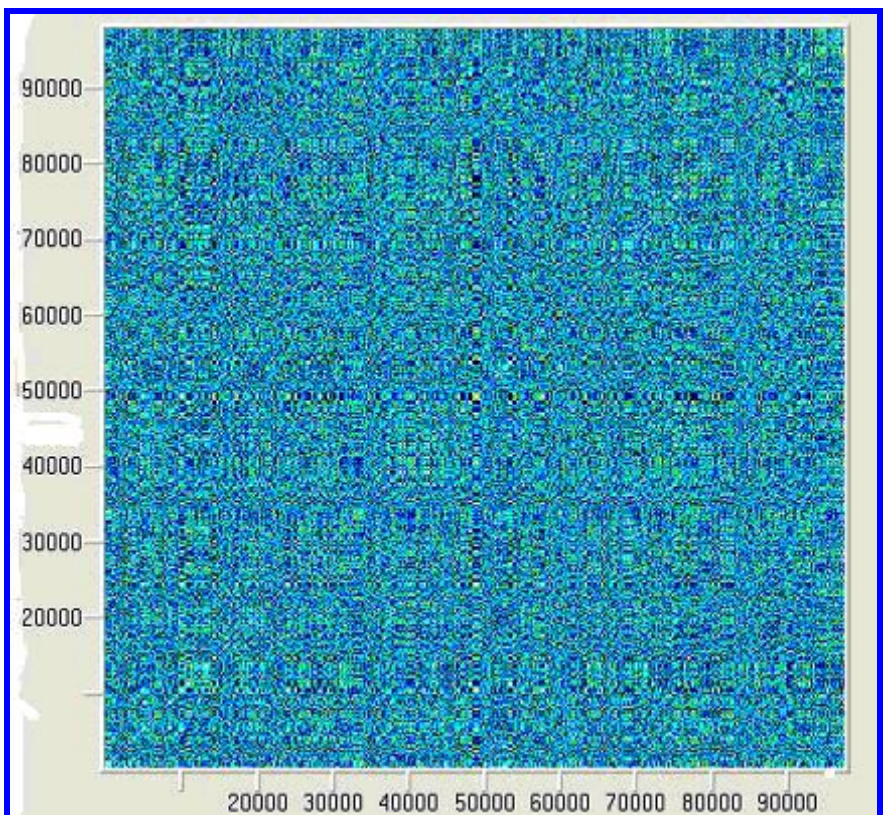

Fig.1f, Recurrence Plot of non-coding HUMTCRADCV (Embed. Dim. D=5)

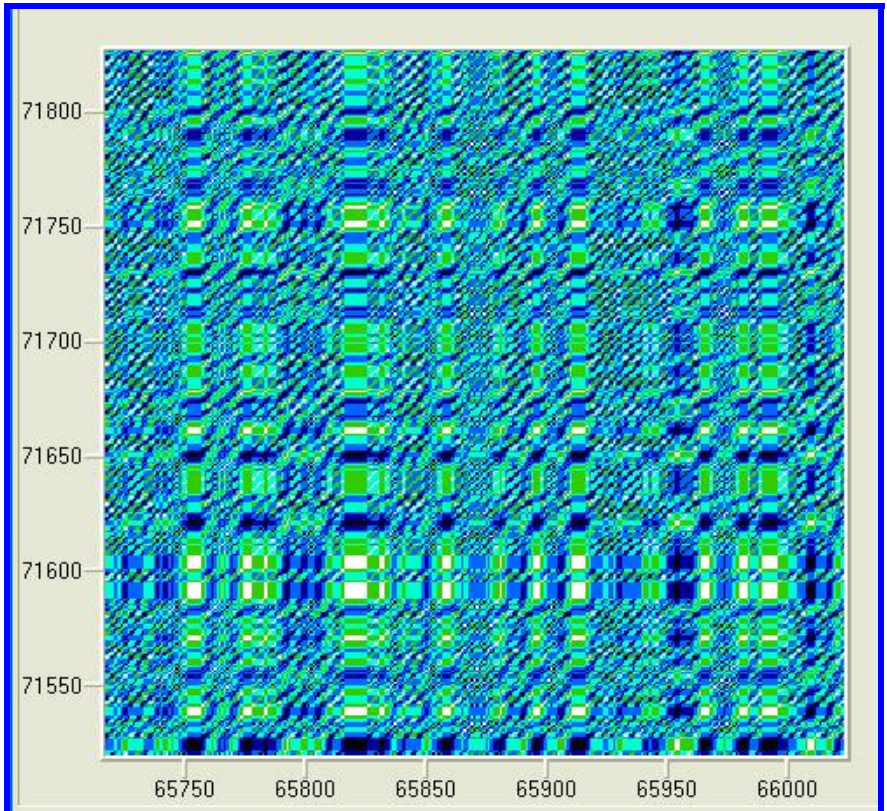
Fig. 1g. Recurrence Plot of non-coding HUMTCRADCV (Embed. Dim. D=5)
It is a selected section of the whole map given in Fig. 1f

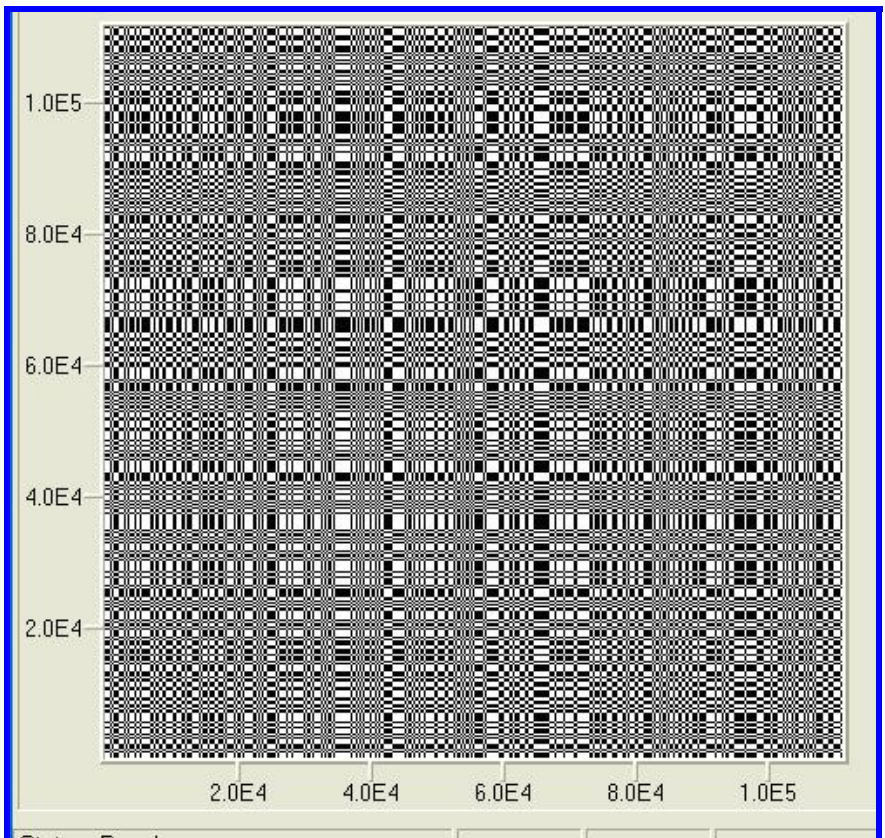
Fig.2a. Recurrence Plot of ECO110K-genomic fragment (Embed. Dim. D=1)

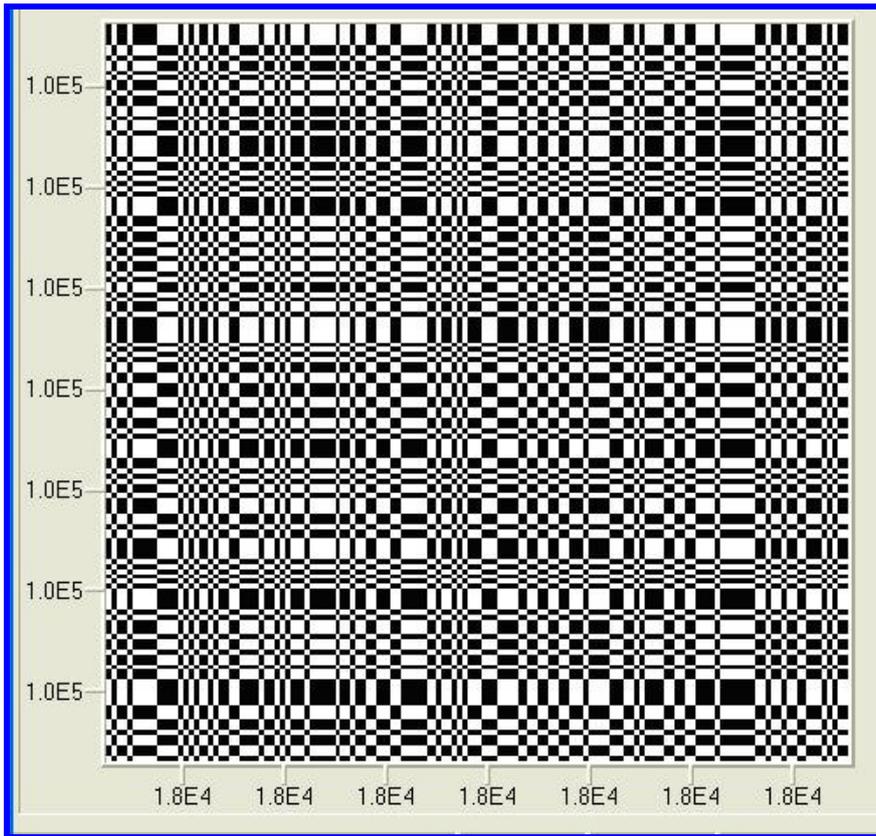
Fig.2b Recurrence Plot of ECO110K-genomic fragment (Embed. Dim. D=1)
It is a selected section of the whole map given in Fig.2a

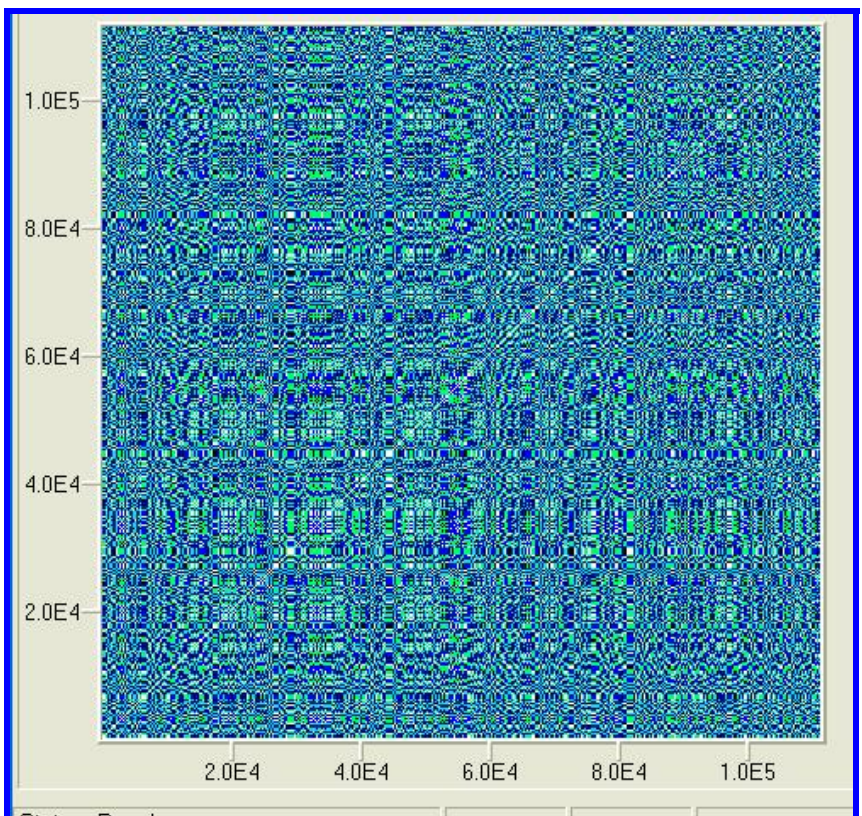
Fig.2c Recurrence Plot of ECO110K-genomic fragment (Embed. Dim. D=3)

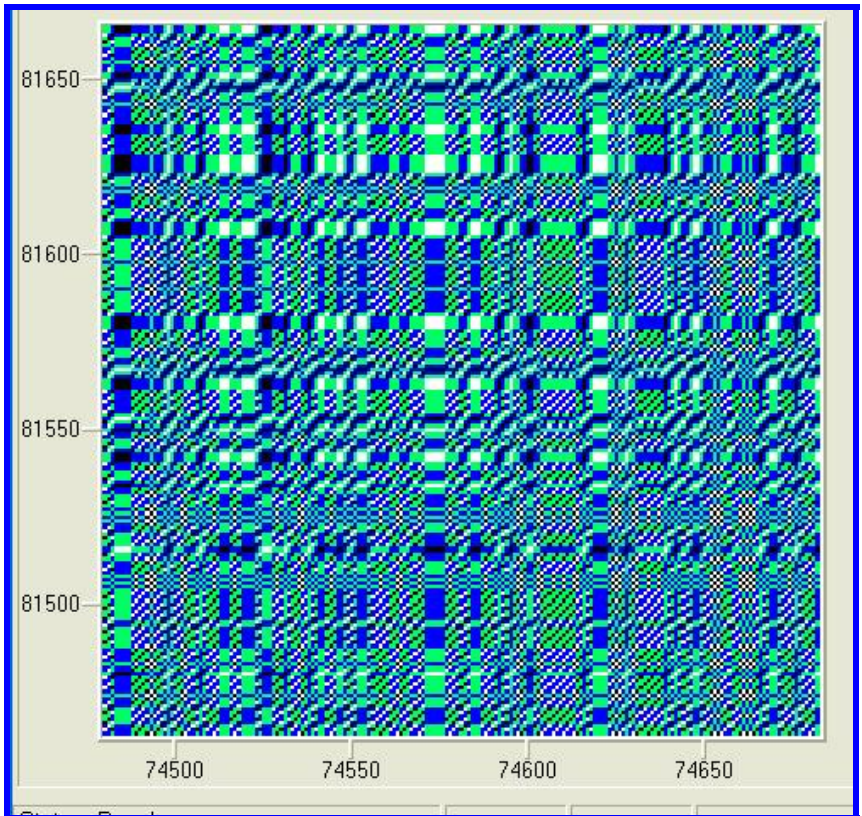

Fig.2d Recurrence Plot of ECO110K-genomic fragment (Embed. Dim. D=3)
It is a selected section of the whole map given in Fig.2c.

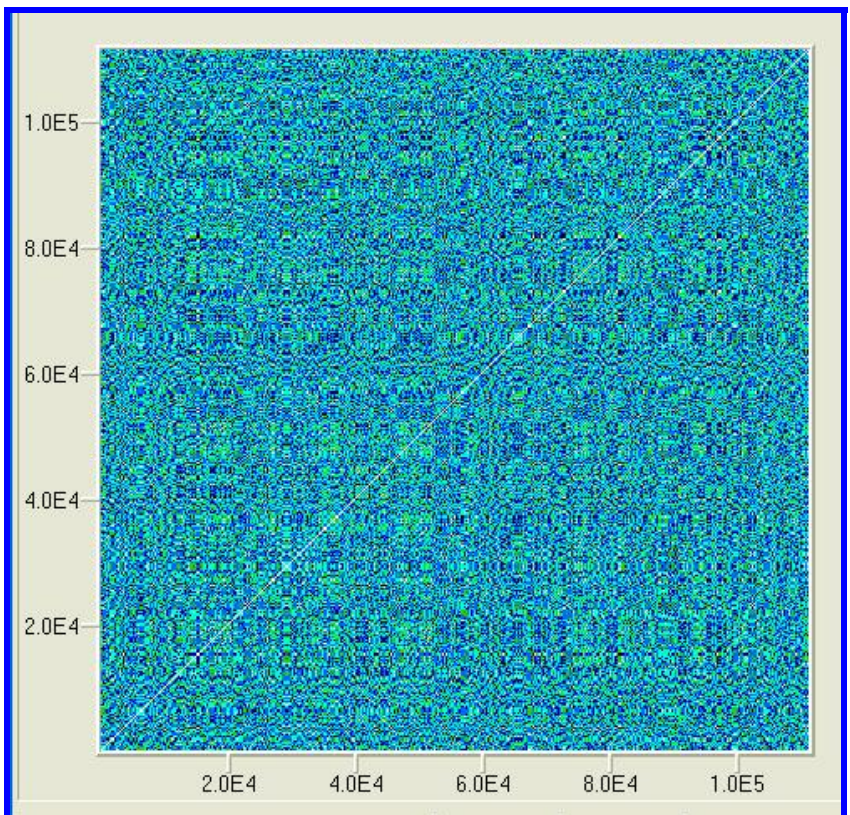

Fig.2e Recurrence Plot of ECO110K-genomic fragment (Embed. Dim. D=5)

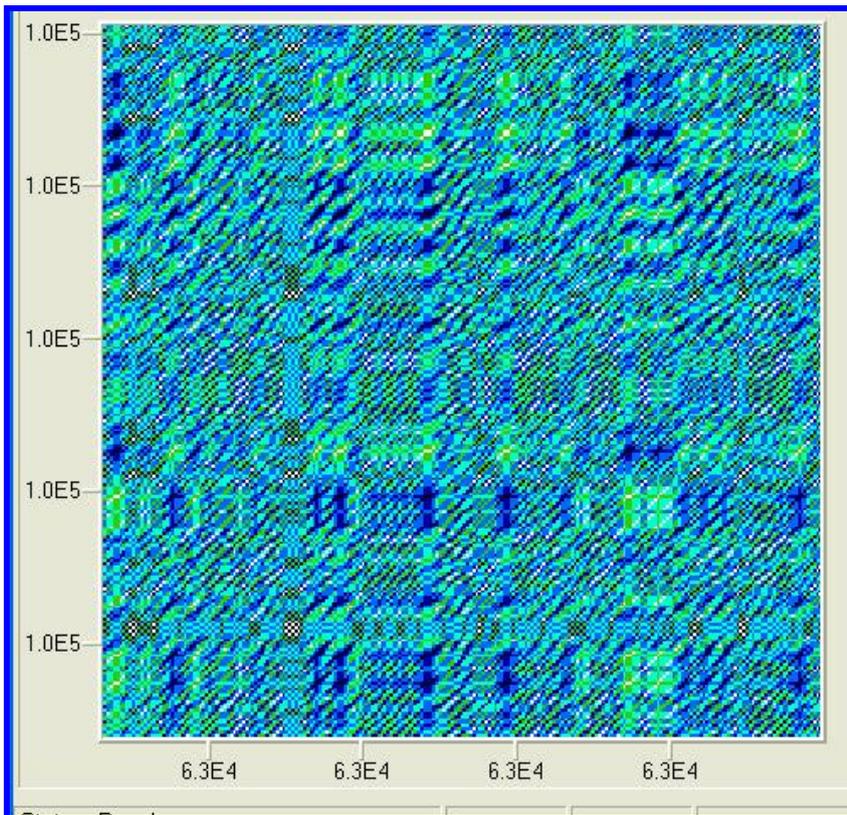
Fig.2f Recurrence Plot of ECO110K-genomic fragment (Embed. Dim. D=5)
It is a selected section of the whole map given in Fig. 2e

# RQD –ANALYSIS

| Sequences | N. Recurrences | N. Lines | %Rec | %Det | L. Max | Entropy | %Laminarity | Ratio |
|---|---|---|---|---|---|---|---|---|
| Non coding HUMTCRADCV (97630 bases) (Embed. Dim. **D=1**) | 6806151 | 868019 | 50.062 | 51.703 | 61 | 2.052 | 60.183 | 1.033 |
| ECO110K-genomic fragment(111401 bases) (Embed. Dim. **D=1**) | 6802929 | 842000 | 50.038 | 49.896 | 26 | 2.031 | 44.419 | 0.997 |
| Non coding HUMTCRADCV (97630 bases) (Embed. Dim. **D=3**) | 1782959 | 227169 | 13.124 | 51.890 | 59 | 2.069 | 26.695 | 3.953 |
| ECO110K-genomic fragment(111401 bases) (Embed. Dim. **D=3**) | 1710381 | 216694 | 12.590 | 51.189 | 24 | 2.040 | 10.347 | 4.066 |
| Non coding HUMTCRADCV (97630 bases) (Embed. Dim. **D=5**) | 470840 | 59994 | 3.469 | 52.472 | 57 | 2.106 | 14.548 | 15.125 |
| ECO110K-genomic fragment(111401 bases) (Embed. Dim. **D=5**) | 442140 | 56319 | 3.257 | 51.502 | 22 | 2.042 | 3.114 | 15.812 |

**Table 1. Results of the RQD analysis for the fragments Non- coding HUMTCRADCV and coding ECO110K**

**RQD –ANALYSIS with Shuffled data**

| Sequences | N. Recurrences | N. Lines | %Rec | %Det | L. Max | Entropy | %Laminarity | Ratio |
|---|---|---|---|---|---|---|---|---|
| Non coding HUMTCRADCV (97630 bases) (Embed. Dim. **D=1**) | 6806151 | 868019 | 50.062 | 51.703 | 61 | 2.052 | 60.183 | 1.033 |
| Shuffled data | 6786451 | 857999 | 49.990 | 50.846 | 23 | 2.027 | 57.644 | 1.017 |
| ECO110K-genomic fragment (111401 bases) (Embed. Dim. **D=1**) | 6802929 | 842000 | 50.038 | 49.896 | 26 | 2.031 | 44.419 | 0.997 |
| Shuffled data | 6797709 | 848780 | 50.000 | 49.996 | 24 | 2.002 | 48.382 | 0.999 |
| Non coding HUMTCRADCV (97630 bases) (Embed. Dim. **D=3**) | 1782959 | 227169 | 13.124 | 51.890 | 59 | 2.069 | 26.695 | 3.953 |
| Shuffled data | 1739717 | 220047 | 12.806 | 50.948 | 21 | 2.028 | 24.843 | 3.978 |
| ECO110K-genomic fragment (111401 bases) (Embed. Dim. **D=3**) | 1710381 | 216694 | 12.590 | 51.189 | 24 | 2.040 | 10.347 | 4.066 |
| Shuffled data | 1698985 | 212746 | 12.506 | 50.071 | 22 | 1.999 | 11.271 | 4.003 |
| Non coding HUMTCRADCV (97630 bases) (Embed. Dim. **D=5**) | 470840 | 59994 | 3.469 | 52.472 | 57 | 2.106 | 14.548 | 15.125 |
| Shuffled data | 446262 | 56571 | 3.287 | 51.136 | 19 | 2.033 | 10.430 | 15.557 |
| ECO110K-genomic fragment (111401 bases) (Embed. Dim. **D=5**) | 442140 | 56319 | 3.257 | 51.502 | 22 | 2.042 | 3.114 | 15.812 |
| Shuffled data | 425205 | 53031 | 3.132 | 49.875 | 20 | 1.999 | 2.611 | 15.924 |

**Table 2.** Comparison of sequences and their shuffled data for HUMTCRADCV and ECO110

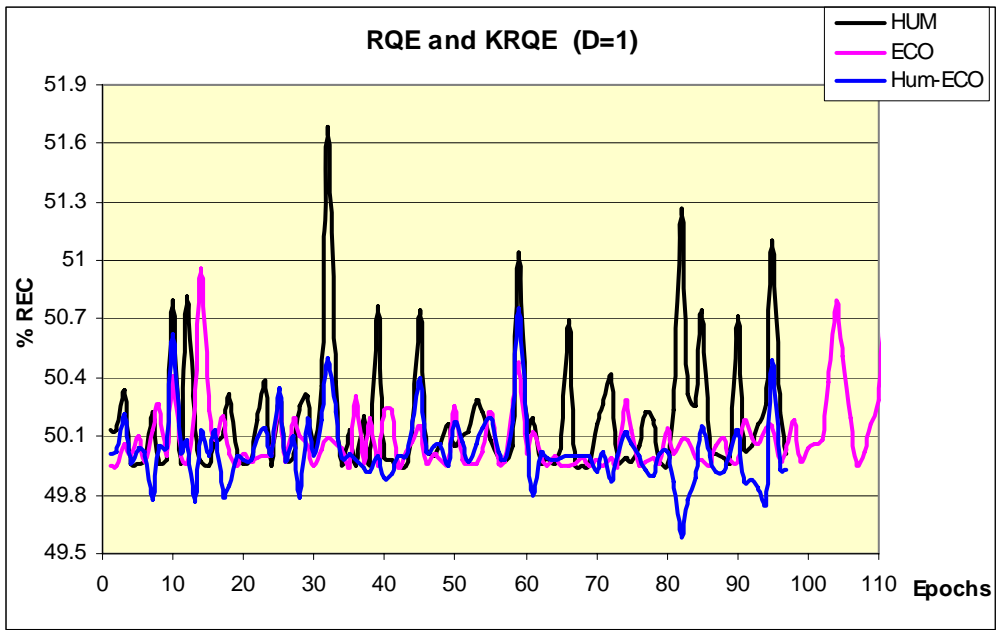

Fig. 3a. Results of RQE and KRQE analysis in the case of non coding HUMTCRADCV and ECO110K-genomic fragment

| | |
|---|---|
| mean value HUMB-d1= | 50.183 |
| st.dev= | 0.317 |
| mean value ECOB-d1= | 50.090 |
| st.dev= | 0.146 |
| mean value HUM-ECO -B-d1= | 50.026 |
| st.dev= | 0.173 |

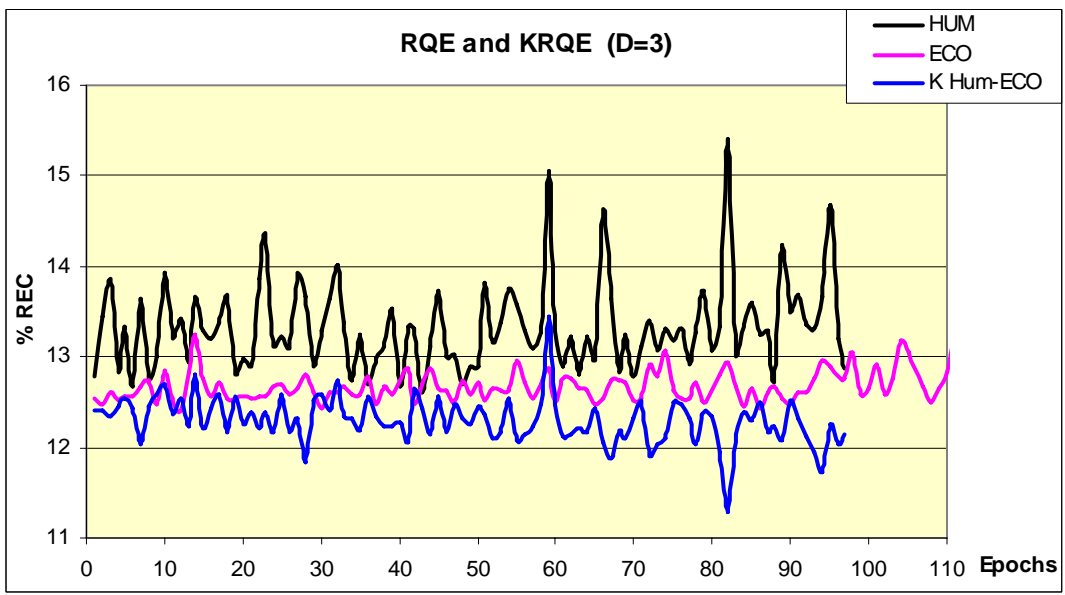

Fig. 3b. Results of RQE and KRQE analysis in the case of non coding HUMTCRADCV and ECO110K-genomic fragment

| | |
|---|---|
| mean value HUMB-d3= | 13.327 |
| st.dev= | 0.498 |
| mean value ECOB-d3= | 12.677 |
| st.dev= | 0.121 |
| mean value HUM-ECO-B-d3= | 12.301 |
| st.dev= | 0.260 |

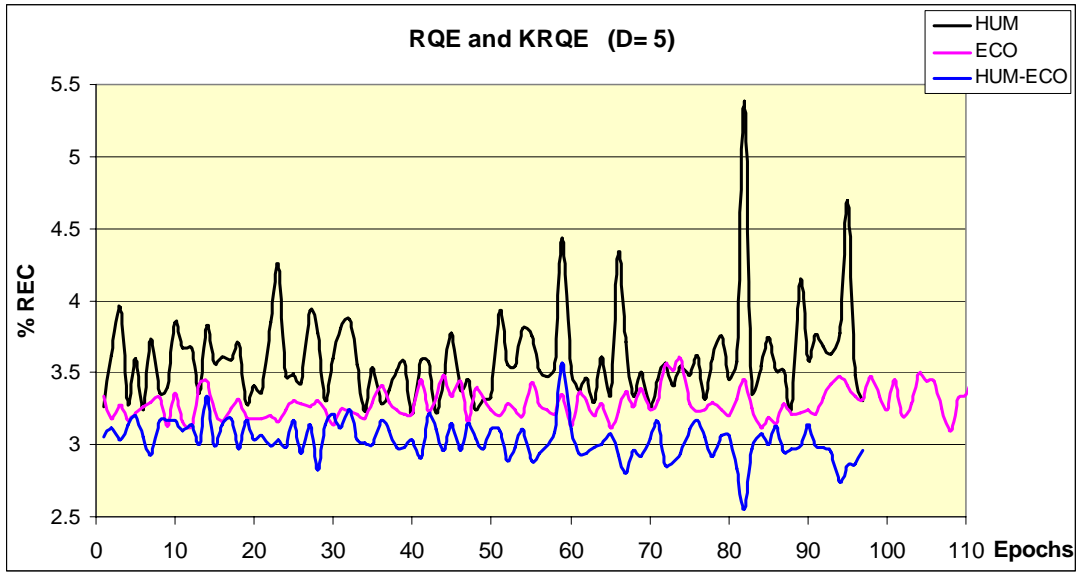

Fig.3c. Results of RQE and KRQE analysis in the case of non coding HUMTCRADCV and ECO110K-genomic fragment

| | |
|---|---|
| mean value HUMB-d5= | 3.590 |
| st.dev= | 0.323 |
| mean value ECOB-d5= | 3.286 |
| st.dev= | 0.105 |
| mean value HUM-ECO-B-d5= | 3.033 |
| st.dev= | 0.130 |

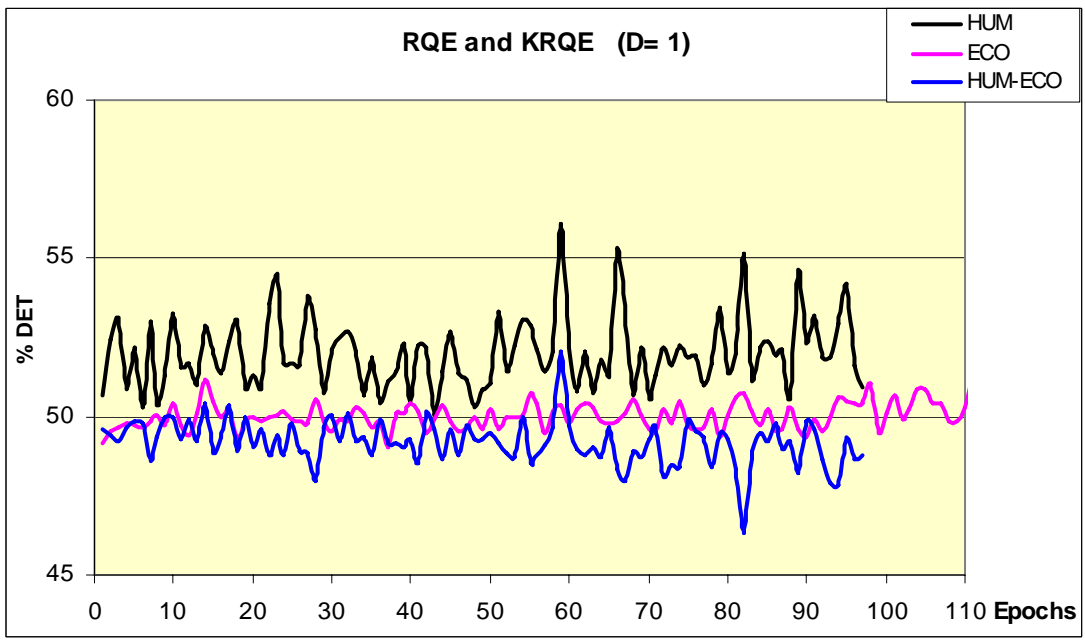

Fig.4a. Results of RQE and KRQE analysis in the case of non coding HUMTCRADCV and ECO110K-genomic fragment

| | |
|---|---|
| mean value HUMB-d1= | 51.990 |
| st.dev= | 1.147 |
| mean value ECOB-d1= | 50.021 |
| st.dev= | 0.421 |
| mean value HUM-ECO-B-d1= | 49.192 |
| st.dev= | 0.701 |

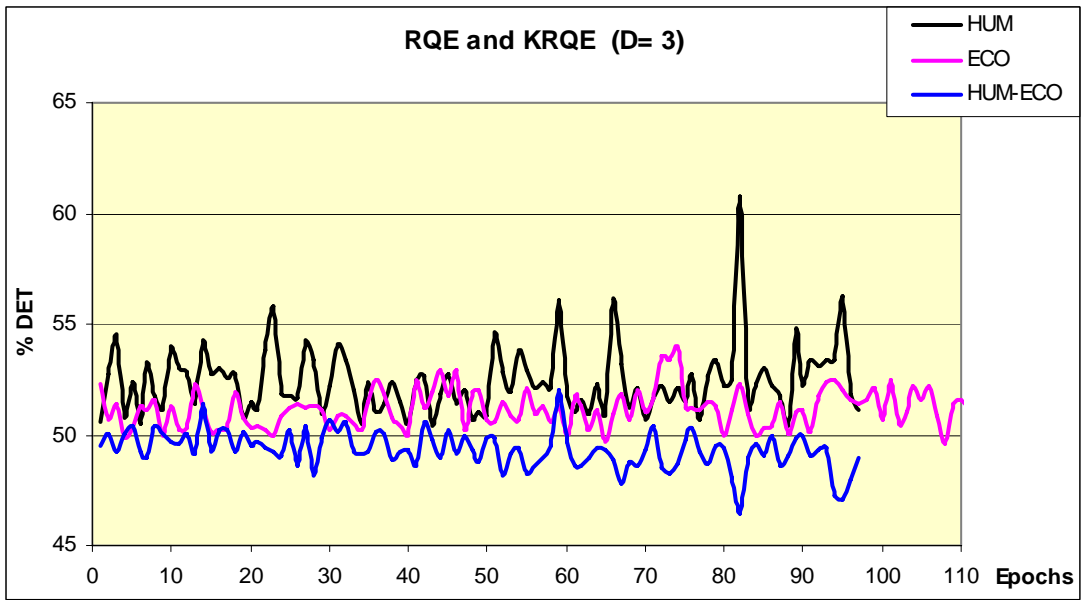

Fig.4b. Results of RQE and KRQE analysis in the case of non coding HUMTCRADCV and ECO110K-genomic fragment

| | |
|---|---|
| mean value HUMB-d3= | 52.409 |
| st.dev= | 1.563 |
| mean value ECOB-d3= | 51.186 |
| st.dev= | 0.864 |
| mean value HUM-ECO-B-d3= | 49.343 |
| st.dev= | 0.853 |

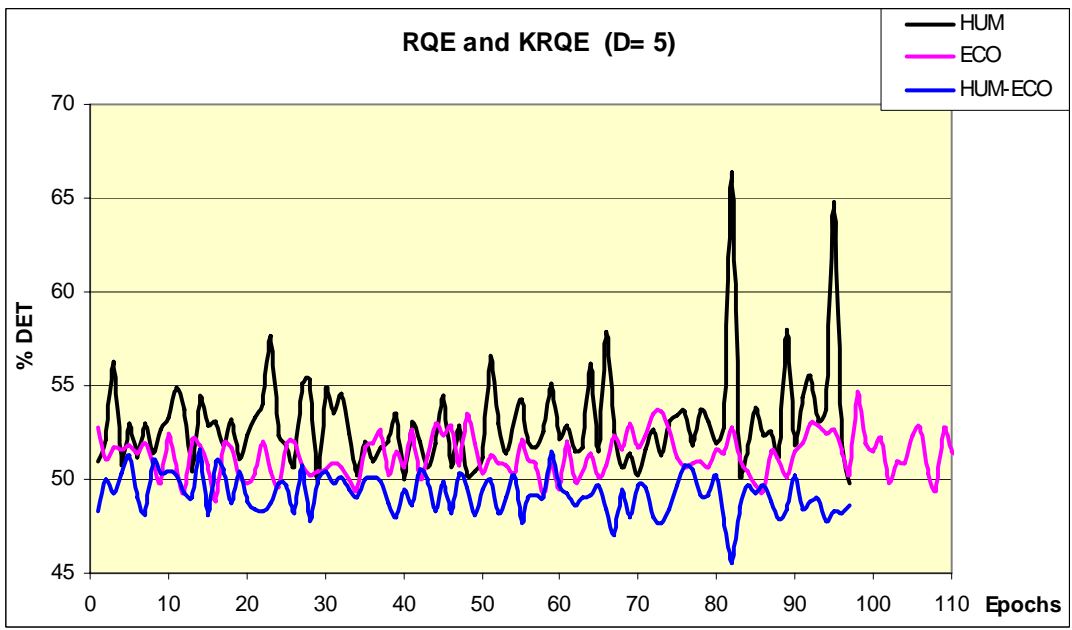

Fig.4c. Results of RQE and KRQE analysis in the case of non coding HUMTCRADCV and ECO110K-genomic fragment

| | |
|---|---|
| mean value HUMB-d5= | 52.927 |
| st.dev= | 2.557 |
| mean value ECOB-d5= | 51.289 |
| st.dev= | 1.120 |
| mean value HUM-ECO-B-d5= | 49.230 |
| st.dev= | 1.045 |

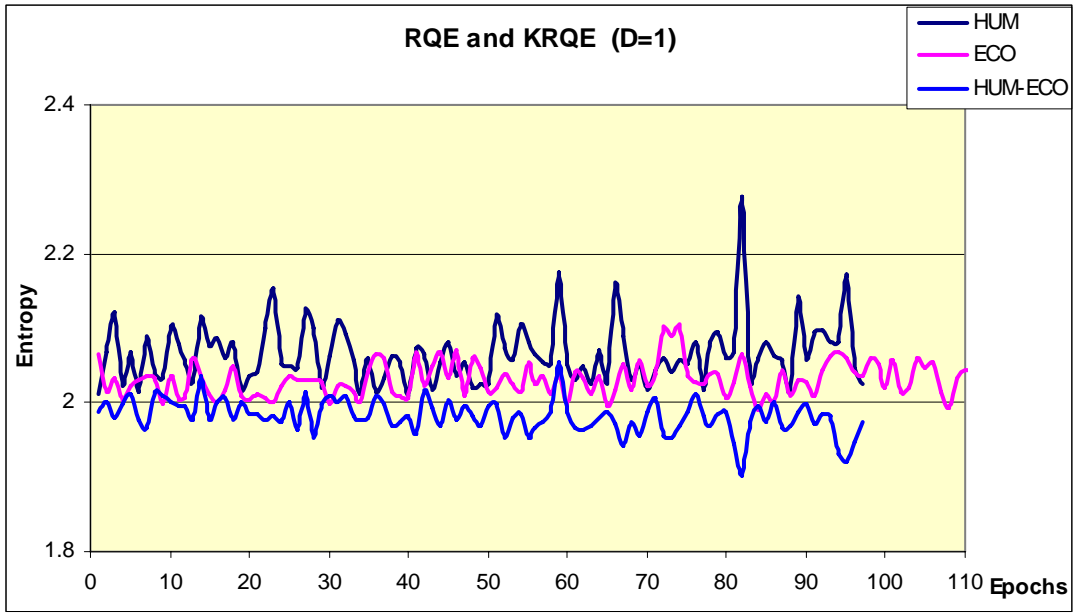

Fig.5a. Results of RQE and KRQE analysis in the case of non coding HUMTCRADCV and ECO110K-genomic fragment

| mean value HUMB-d1= | 2.065 |
|---|---|
| st.dev= | 0.042 |
| mean value ECOB-d1= | 2.031 |
| st.dev= | 0.023 |
| mean value HUM-ECO-B-d1= | 1.982 |
| st.dev= | 0.023 |

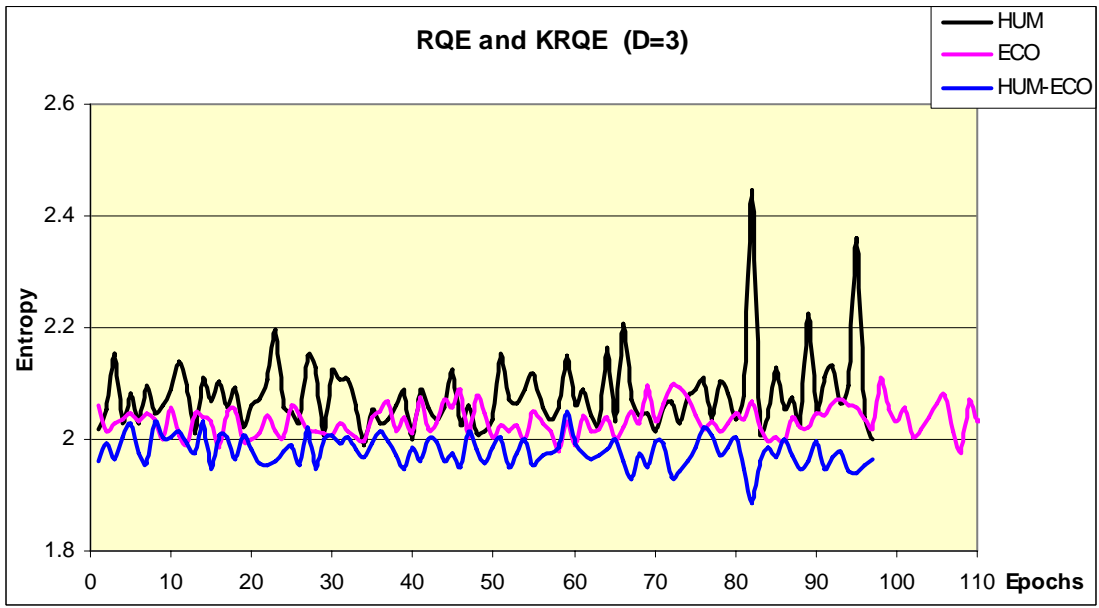

Fig. 5b. Results of RQE and KRQE analysis in the case of non coding HUMTCRADCV and ECO110K-genomic fragment

| mean value HUMB-d3= | 2.078 |
|---|---|
| st.dev= | 0.067 |
| mean value ECOB-d3= | 2.034 |
| st.dev= | 0.027 |
| mean value HUM-ECO-B-d3= | 1.978 |
| st.dev= | 0.027 |

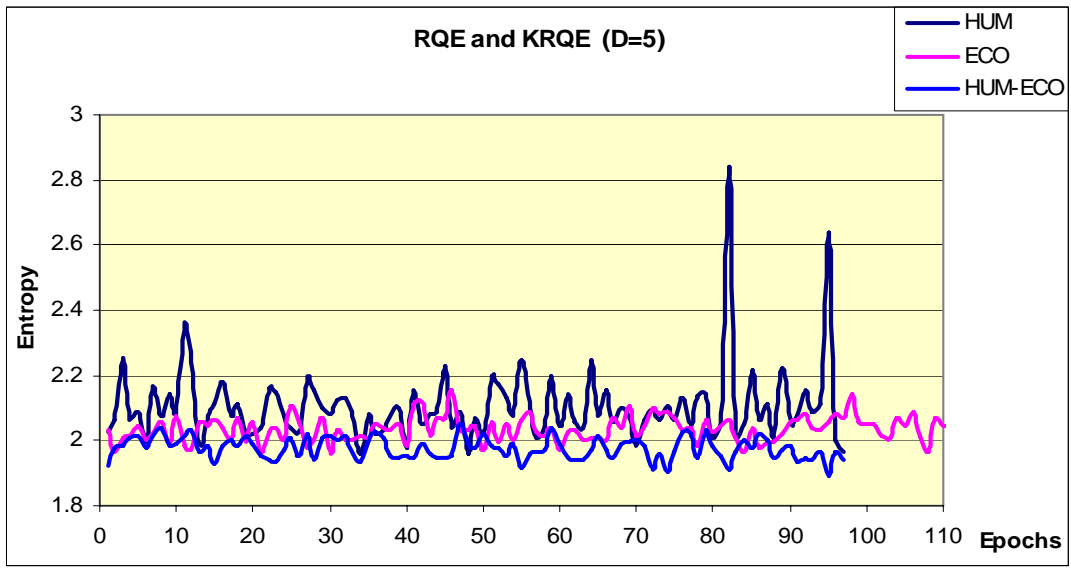

Fig. 5c. Results of RQE and KRQE analysis in the case of non coding HUMTCRADCV and ECO110K-genomic fragment

| | |
|---|---|
| mean value HUMB-d5= | 2.106 |
| st.dev= | 0.119 |
| mean value ECOB-d5= | 2.035 |
| st.dev= | 0.039 |
| mean value HUM-ECO-B-d5= | 1.976 |
| st.dev= | 0.033 |

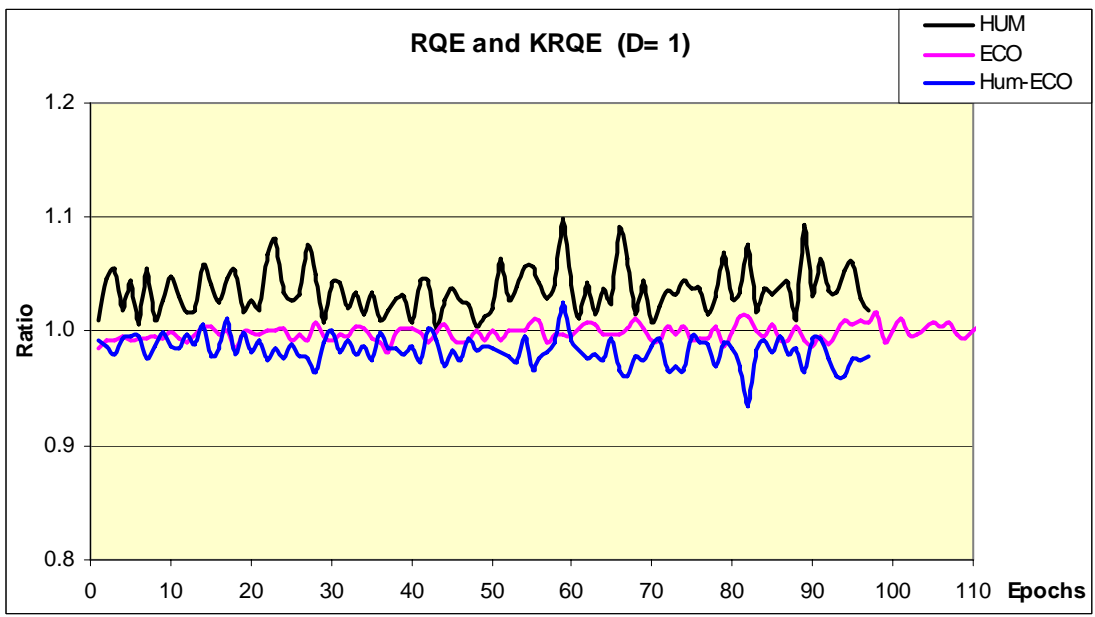

Fig.6a Results of RQE and KRQE analysis in the case of non coding HUMTCRADCV and ECO110K-genomic fragment

| | |
|---|---|
| mean value HUMB-d1= | 1.036 |
| st.dev= | 0.020 |
| mean value ECOB-d1= | 0.999 |
| st.dev= | 0.007 |
| mean value HUM-ECO-B-d1= | 0.983 |
| st.dev= | 0.012 |

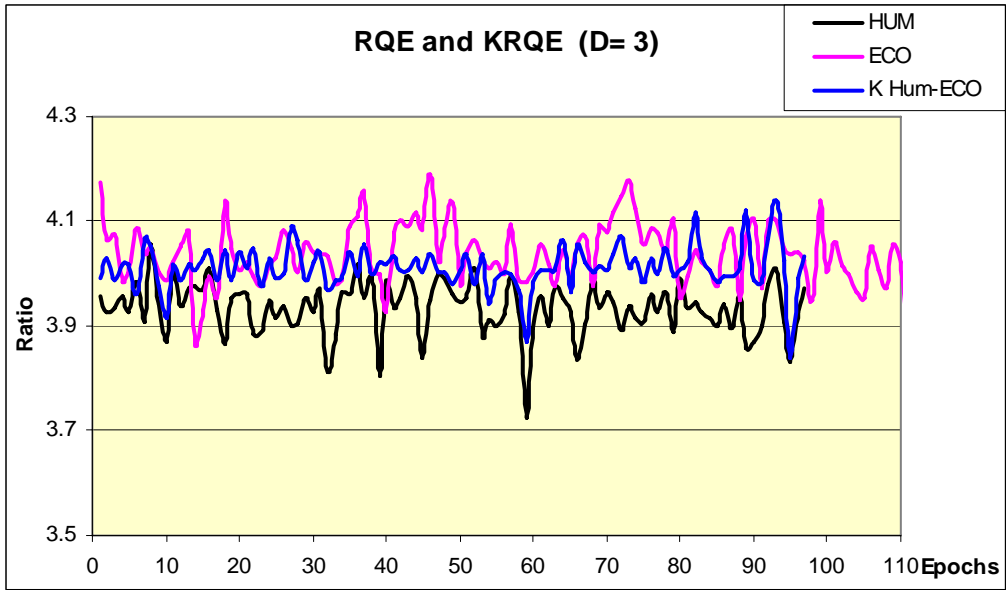

Fig. 6b. Results of RQE and KRQE analysis in the case of non coding HUMTCRADCV and ECO110K-genomic fragment

| | |
|---|---|
| mean value HUMB-d3= | 3.934 |
| st.dev= | 0.051 |
| mean value ECOB-d3= | 4.038 |
| st.dev= | 0.060 |
| mean value HUM-ECO-B-d3= | 4.012 |
| st.dev= | 0.042 |

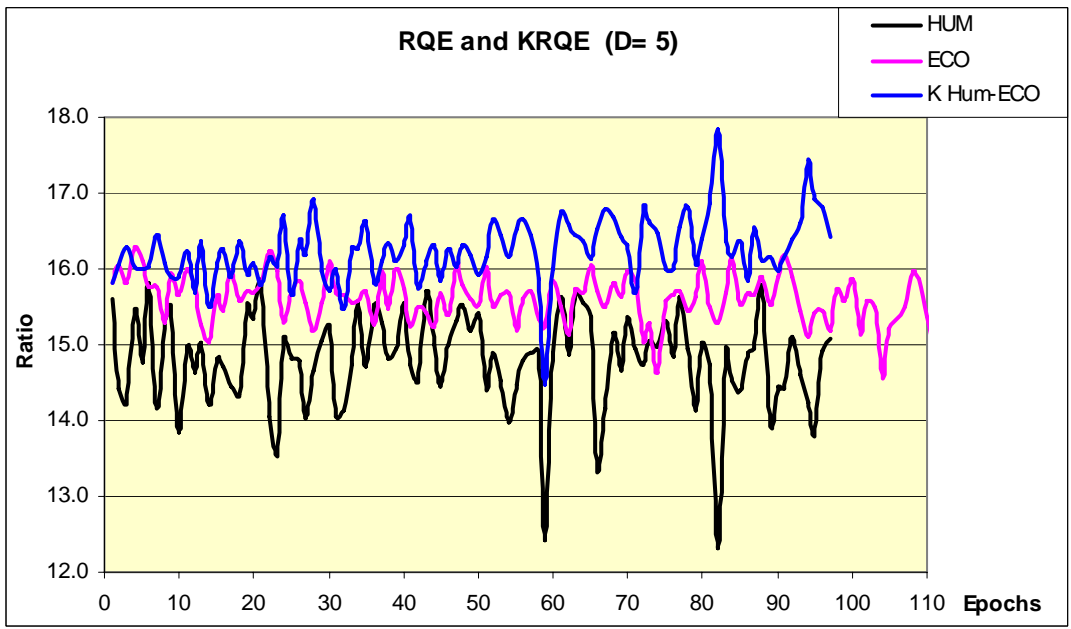

Fig. 6c. Results of RQE and KRQE analysis in the case of non coding HUMTCRADCV and ECO110K-genomic fragment

| | |
|---|---|
| mean value HUMB-d5= | 14.794 |
| st.dev= | 0.640 |
| mean value ECOB-d5= | 15.614 |
| st.dev= | 0.320 |
| mean value HUM-ECO-B-d5= | 16.248 |
| st.dev= | 0.425 |

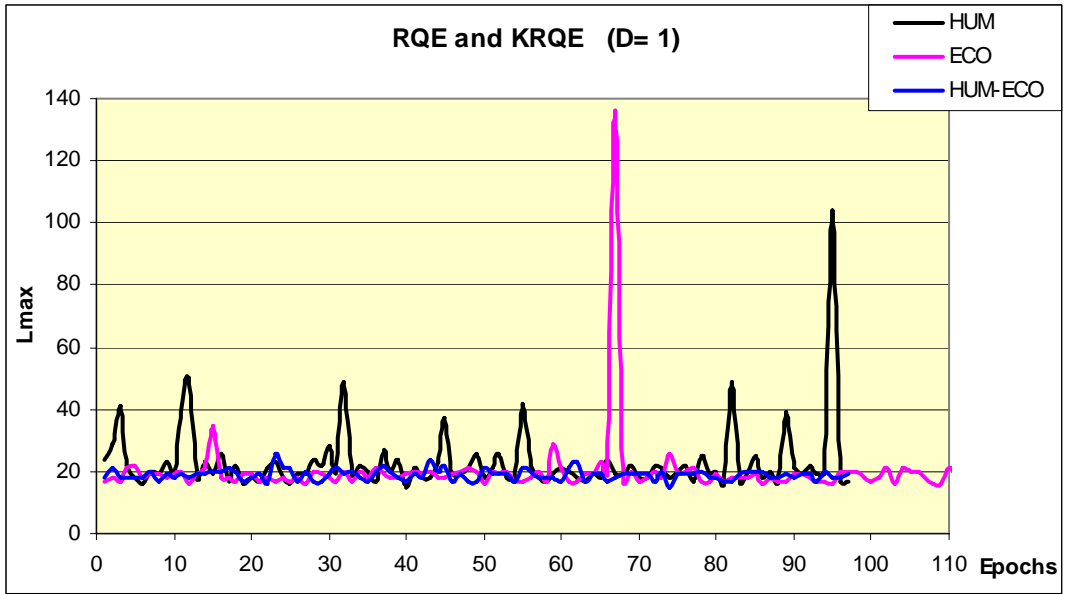

Fig. 7a Results of RQE and KRQE analysis in the case of non coding HUMTCRADCV and ECO110K-genomic fragment

| | |
|---|---|
| mean value HUMB-d1= | 22.845 |
| st.dev= | 10.965 |
| mean value ECOB-d1= | 19.892 |
| st.dev= | 11.392 |
| mean value HUM-ECO-B-d1= | 18.887 |
| st.dev= | 1.689 |

.

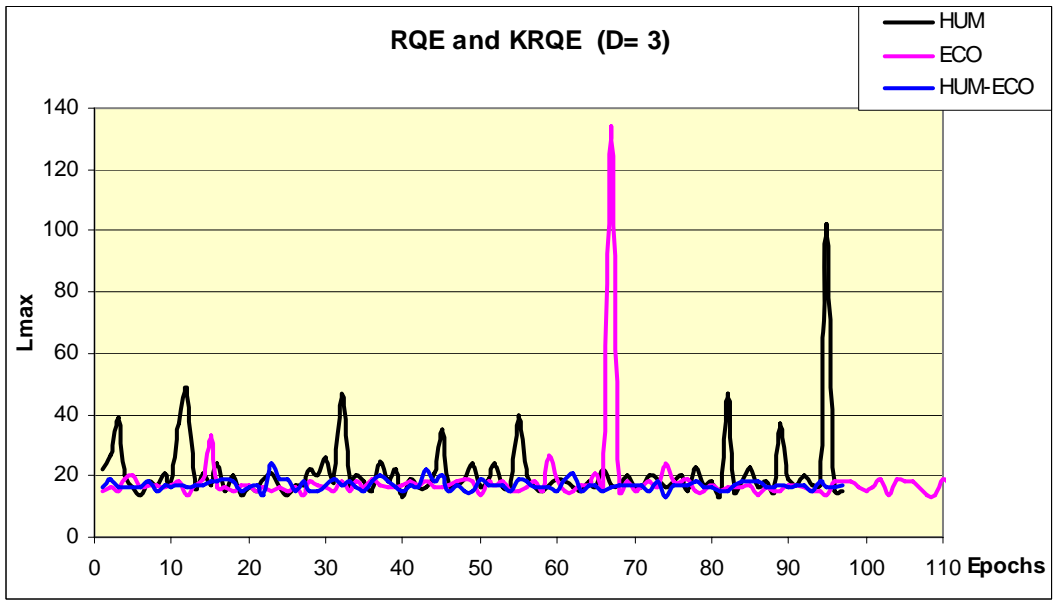

Fig. 7b Results of RQE and KRQE analysis in the case of non coding HUMTCRADCV and ECO110K-genomic fragment

| | |
|---|---|
| mean value HUMB-d3= | 20.856 |
| st.dev= | 10.962 |
| mean value ECOB-d3= | 17.892 |
| st.dev= | 11.392 |
| mean value HUM-ECO-B-d3= | 16.887 |
| st.dev= | 1.689 |

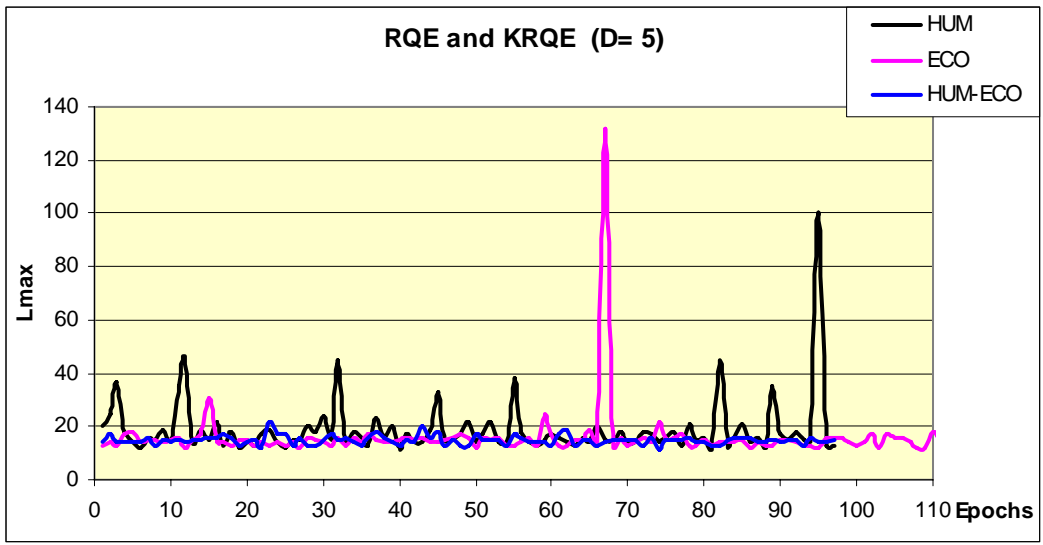

Fig. 7c Results of RQE and KRQE analysis in the case of non coding HUMTCRADCV and ECO110K-genomic fragment

| | |
|---|---|
| mean value HUMB-d5= | 18.856 |
| st.dev= | 10.962 |
| mean value ECOB-d5= | 15.901 |
| st.dev= | 11.394 |
| mean value HUM-ECO-B-d5= | 14.887 |
| st.dev= | 1.689 |

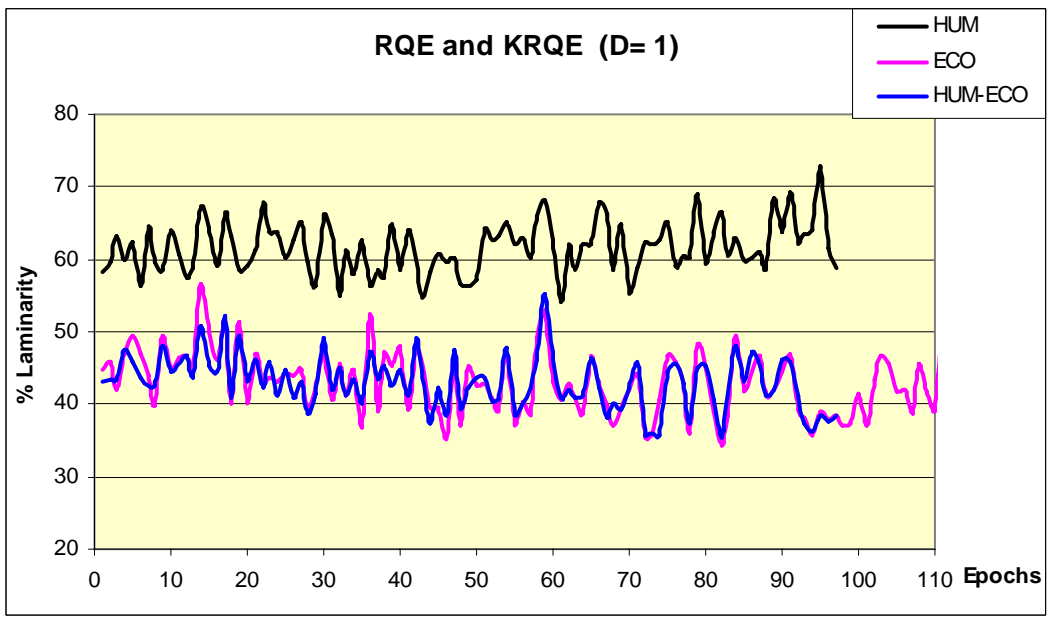

Fig. 8a Results of RQE and KRQE analysis in the case of non coding HUMTCRADCV and ECO110K-genomic fragment

| | |
|---|---|
| mean value HUMB-d1= | 61.621 |
| st.dev= | 3.611 |
| mean value ECOB-d1= | 43.027 |
| st.dev= | 4.455 |
| mean value HUM-ECO-B-d1= | 43.038 |
| st.dev= | 3.740 |

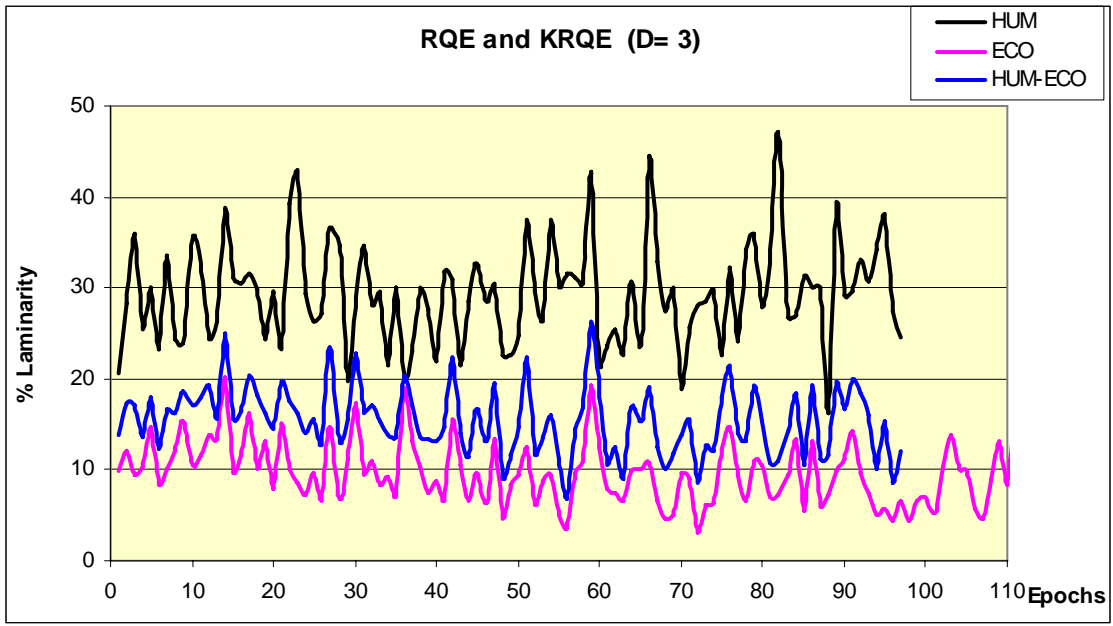

Fig. 8b Results of RQE and KRQE analysis in the case of non coding HUMTCRADCV and ECO110K-genomic fragment

| | |
|---|---|
| mean value HUMB-d3= | 29.306 |
| st.dev= | 5.766 |
| mean value ECOB-d3= | 9.717 |
| st.dev= | 3.551 |
| mean value HUM-ECO-B-d3= | 15.422 |
| st.dev= | 3.771 |

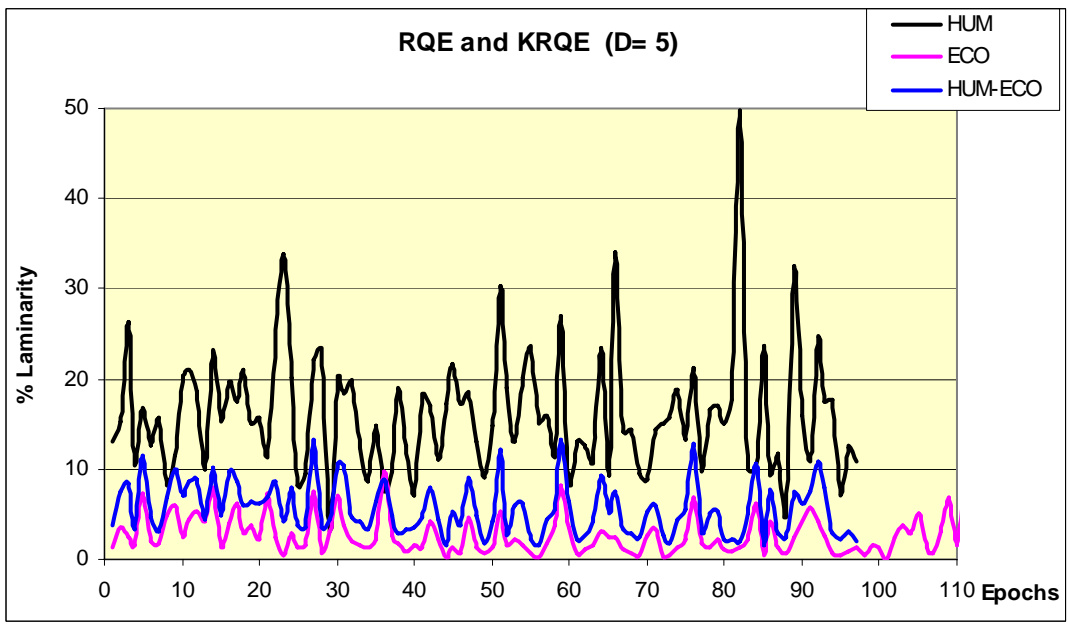

Fig. 8c Results of RQE and KRQE analysis in the case of non coding HUMTCRADCV and ECO110K-genomic fragment

| | |
|---|---|
| mean value HUMB-d5= | 16.179 |
| st.dev= | 7.017 |
| mean value ECOB-d5= | 2.715 |
| st.dev= | 2.172 |
| mean value HUM-ECO-B-d5= | 5.629 |
| st.dev= | 2.972 |

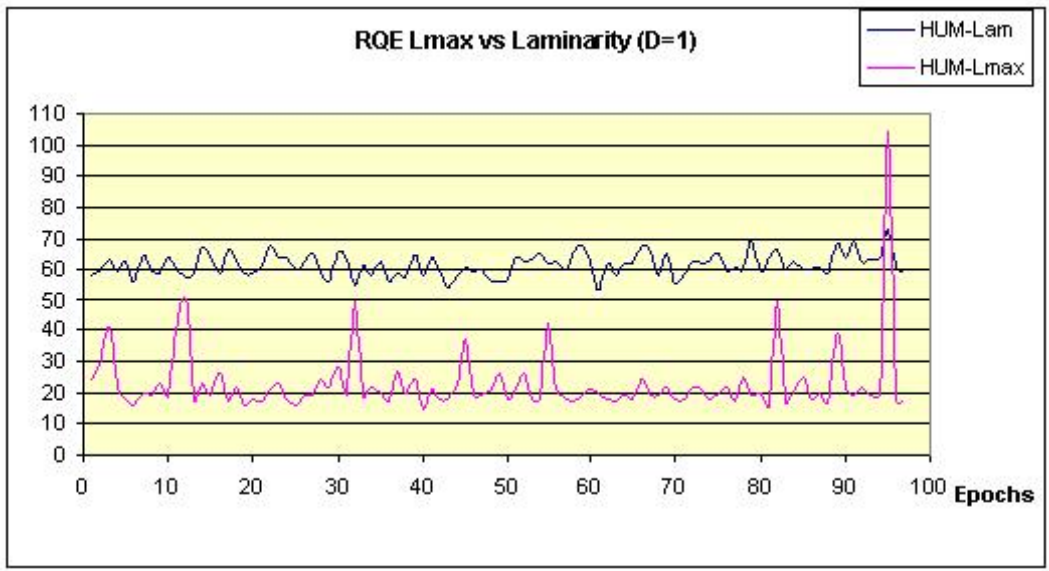

Fig. 9a. Comparison of %Laminarity vs Lmax in HUMTCRADCV with Embed Dim D=1

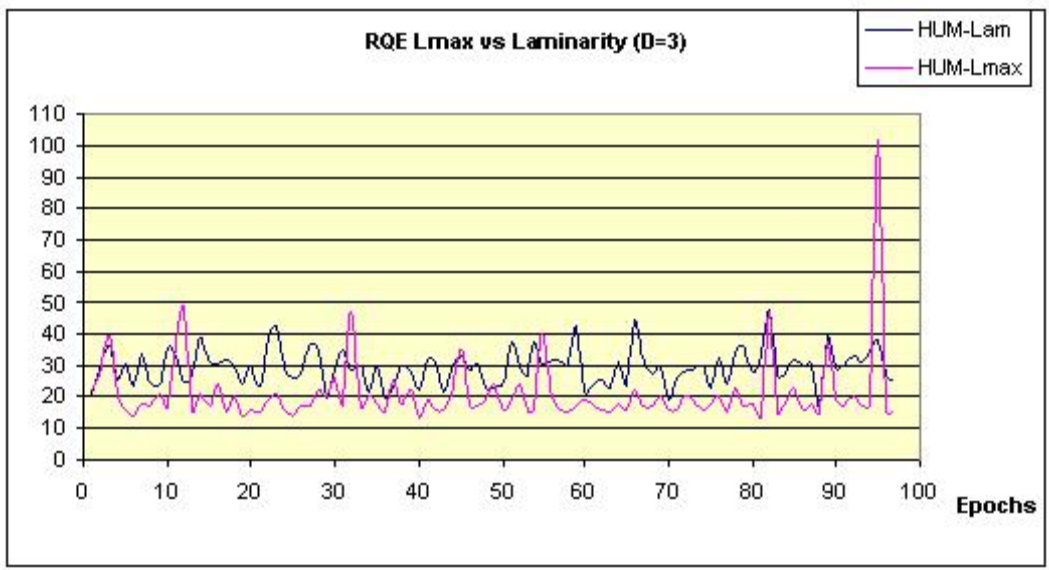

Fig. 9b. Comparison of %Laminarity vs Lmax in HUMTCRADCV with Embed Dim D=3

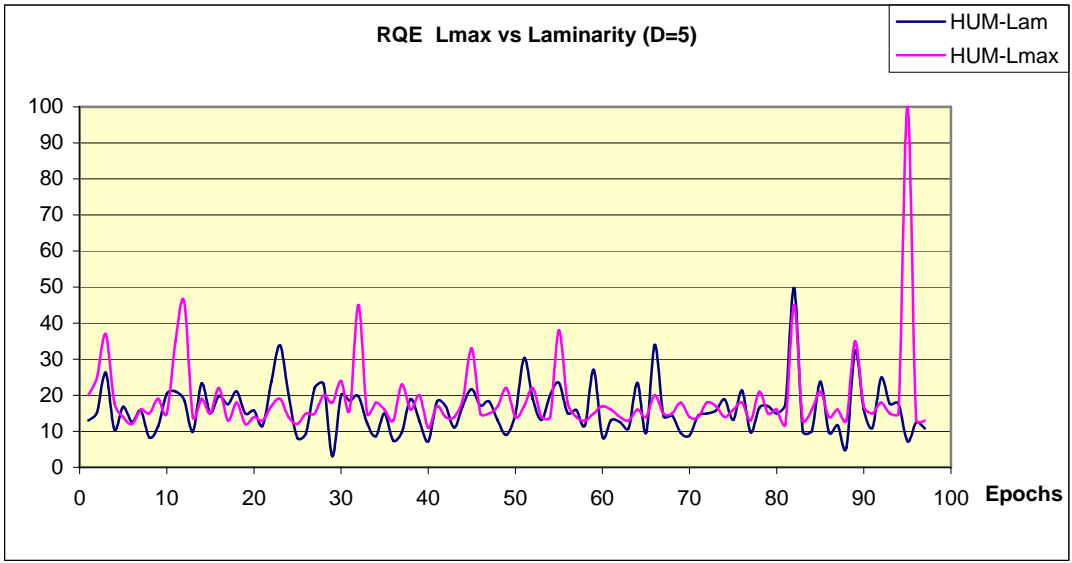

Fig. 9c. Comparison of %Laminarity vs Lmax in HUMTCRADCV with Embed Dim D=5

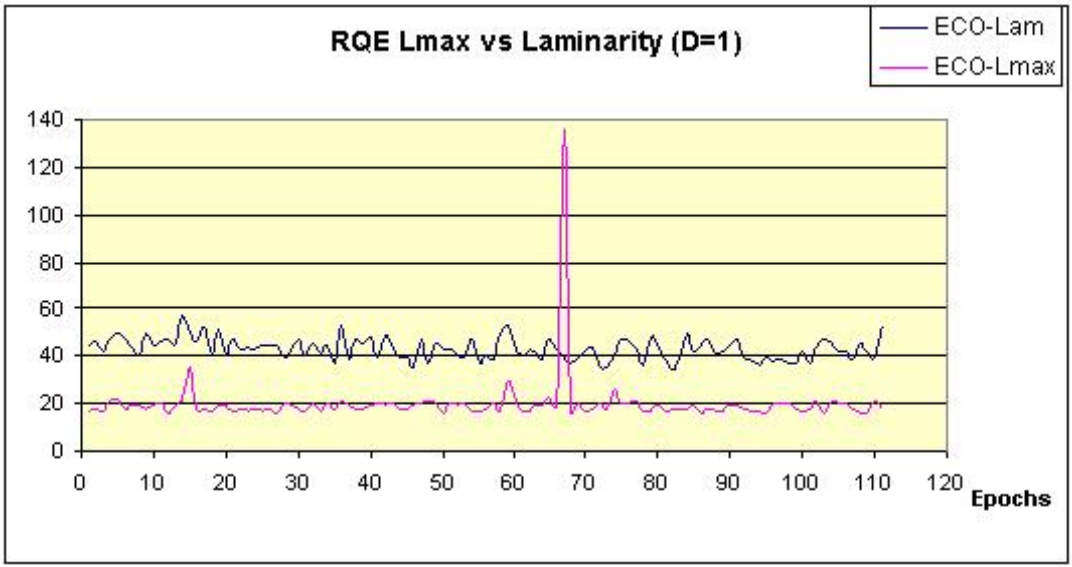

Fig.10a. Comparison of %Laminarity vs Lmax in ECO110K with Embed Dim D=1

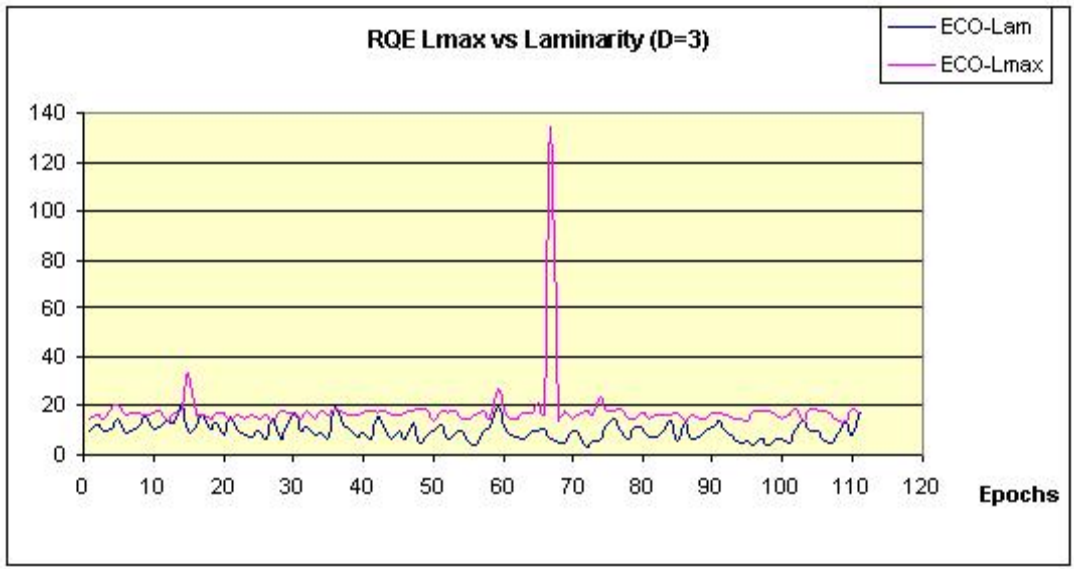

Fig. 10b. Comparison of %Laminarity vs Lmax in ECO110K with Embed Dim D=3

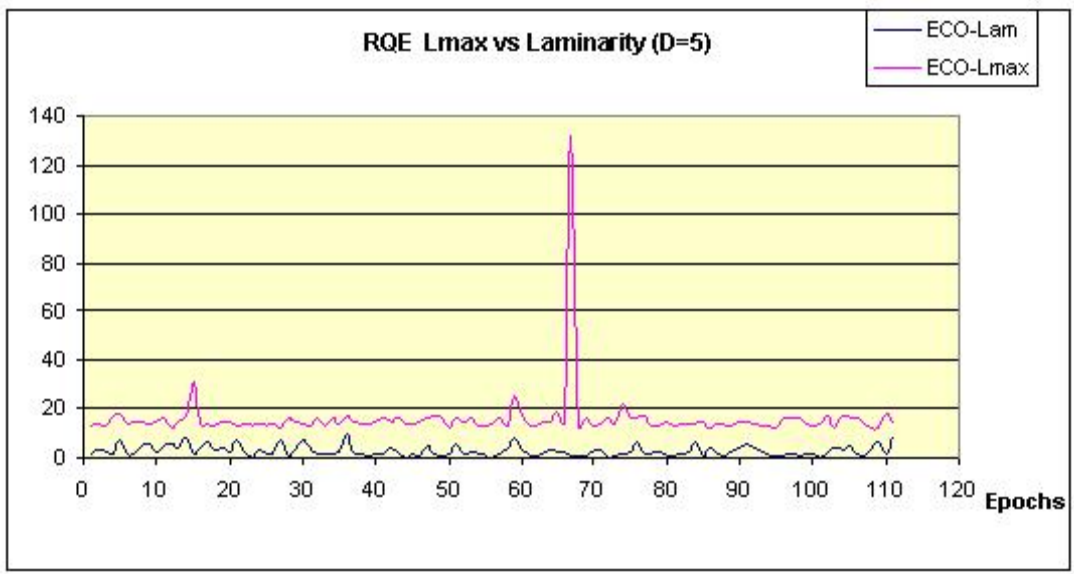

Fig.10c. Comparison of %Laminarity vs Lmax in ECO110K with Embed Dim D=5